\documentclass[conference]{IEEEtran}
\IEEEoverridecommandlockouts

\def\BibTeX{{\rm B\kern-.05em{\sc i\kern-.025em b}\kern-.08em
    T\kern-.1667em\lower.7ex\hbox{E}\kern-.125emX}}

\usepackage{cite}
\usepackage{amsmath,amssymb,amsfonts}
\usepackage{graphicx}
\usepackage{textcomp}
\usepackage{xcolor}
\usepackage{braket}
\usepackage{comment}
\usepackage{mathtools}
\usepackage[most]{tcolorbox}
\usepackage{booktabs}   
\usepackage{array}      
\usepackage{multirow}   
\usepackage{subcaption}
\usepackage{hyperref}

\newcommand{\etal}{\textit{et al}.}

\newtcolorbox{rqsummary}{
  enhanced,
  breakable,
  colback=gray!10,   
  colframe=black!70, 
  boxrule=1pt,       
  arc=1mm,           
  left=1mm, right=1mm, top=1mm, bottom=1mm
}

\makeatletter
\renewcommand*\env@matrix[1][*\c@MaxMatrixCols c]{
  \hskip -\arraycolsep
  \let\@ifnextchar\new@ifnextchar
  \array{#1}}
\makeatother

\begin{document}
\title{Leveraging Phase Polynomials for Quantum Circuit Optimization}

\author{

\IEEEauthorblockN{Zihan Chen}
\IEEEauthorblockA{
Rutgers University\\
Piscataway, NJ, USA\\
zihan.chen.cs@rutgers.edu
}
\vspace{0.5cm}
\IEEEauthorblockN{Mingkuan Xu}
\IEEEauthorblockA{
Carnegie Mellon University\\
Pittsburgh, PA, USA\\
mingkuan@cmu.edu
}
\vspace{-0.5cm}

\and
\IEEEauthorblockN{Henry Chen}
\IEEEauthorblockA{
Rutgers University\\
Piscataway, NJ, USA\\
hc867@rutgers.edu
}
\vspace{0.5cm}
\IEEEauthorblockN{Vannessa Chan}
\IEEEauthorblockA{
Rutgers University\\
Piscataway, NJ, USA\\
vlc74@rutgers.edu
}
\vspace{-0.5cm}

\and
\IEEEauthorblockN{Yuwei Jin}
\IEEEauthorblockA{
Rutgers University\\
Piscataway, NJ, USA\\
jyw413482880@gmail.com
}
\vspace{0.5cm}
\IEEEauthorblockN{Won Woo Ro}
\IEEEauthorblockA{
Yonsei University\\
Seoul, Korea\\
wro@yonsei.ac.kr
}
\vspace{-0.5cm}

\and
\IEEEauthorblockN{Enhyeok Jang}
\IEEEauthorblockA{
Yonsei University\\
Seoul, Korea\\
enhyeok.jang@yonsei.ac.kr
}
\vspace{0.5cm}
\IEEEauthorblockN{Eddy Z. Zhang}
\IEEEauthorblockA{
Rutgers University\\
Piscataway, NJ, USA\\
eddy.zhengzhang@gmail.com
}
\vspace{-0.5cm}
}

\maketitle

\begin{abstract}

    Quantum circuits on resource-limited hardware require optimizing regions dominated by $\{\mathrm{CNOT}, R_z\}$, which account for a large fraction of operations and often dominate execution cost.
    This optimization can be challenging because phase-polynomial blocks are fragmented by basis-changing gates such as $H$, and optimizing phase parities alone may increase the cost of downstream basis transformations.
    Existing phase-polynomial approaches are limited to single-block or phase-only optimization, while subcircuit rewriting approaches are local and scale poorly beyond small rewrite windows.
    We introduce \emph{PhasePoly}, a compiler optimization pass that jointly optimizes phase-parity and output-parity networks and employs a cross-block intermediate representation to reuse parities across phase-polynomial block barriers.
    This approach is effective because its unified parity-matrix representation exposes long-range $\{\mathrm{CNOT}, R_z\}$ structure that local rewriting and single-block methods cannot capture.
    \emph{PhasePoly} reduces total gate count by up to 50.00\% (34.70\% on average) and CNOT count by up to 48.57\% (26.83\% on average), while scaling to large circuits and improving both fault-tolerant compilation and near-term hardware execution.
    \emph{PhasePoly} is available at \href{https://github.com/ruadapt/PhasePoly}{https://github.com/ruadapt/PhasePoly}.

\end{abstract}


\section{Introduction}

Quantum computing has gained increasing attention for its potential to address problems that are intractable for classical computers, including integer factorization~\cite{shor:focs94}, discrete logarithms~\cite{proos2003shor}, database search~\cite{grover1996quantum}, and simulations in physics and chemistry~\cite{feynman2018simulating, cao2019quantum}. However, current quantum hardware remains constrained by limited time-space volume, making efficient algorithm design and program optimization essential.

Circuit optimization is a technique that transforms an input circuit into a semantically equivalent but more efficient form. Its primary goal is to reduce gate count and/or circuit depth, thereby lowering error rates, shortening execution time, and improving fidelity. Circuit optimization is an important part of industrial-strength software frameworks, such as in Qiskit~\cite{gadi_aleksandrowicz_2019_zenodo}, Quilc~\cite{smith2020open}, and TKET~\cite{sivarajah_2020_QuantumScienceTechnology}. 

Our work targets \emph{\underline{phase polynomial optimization}}, a class of transformations over circuits composed of CNOT and $R_z$ gates. In such circuits, CNOT gates compute XOR parities of input qubits, while $R_z$ gates apply phase rotations conditioned on these parities.
The notion of \emph{phase polynomials} was introduced by Amy \etal~\cite{amy2014polynomial} in the \emph{sum-over-paths} form (see Section~\ref{sec:mot}), an efficient intermediate representation (IR) that expresses CNOT $+ R_z$ circuits as parity-controlled $R_z$ rotations followed by output basis transformations. This representation has enabled logical circuit optimization~\cite{amy2018controlled, nam2018npj}, equivalence checking~\cite{amy2017verified,amy2018towards}, and hardware-aware circuit synthesis~\cite{de2020architecture, vandaele2022QuantumScienceTechnology, li2025hopps}.

\begin{figure}[t!]
    \vspace{0.5\baselineskip}
    \centering
    \includegraphics[width=0.45\textwidth]{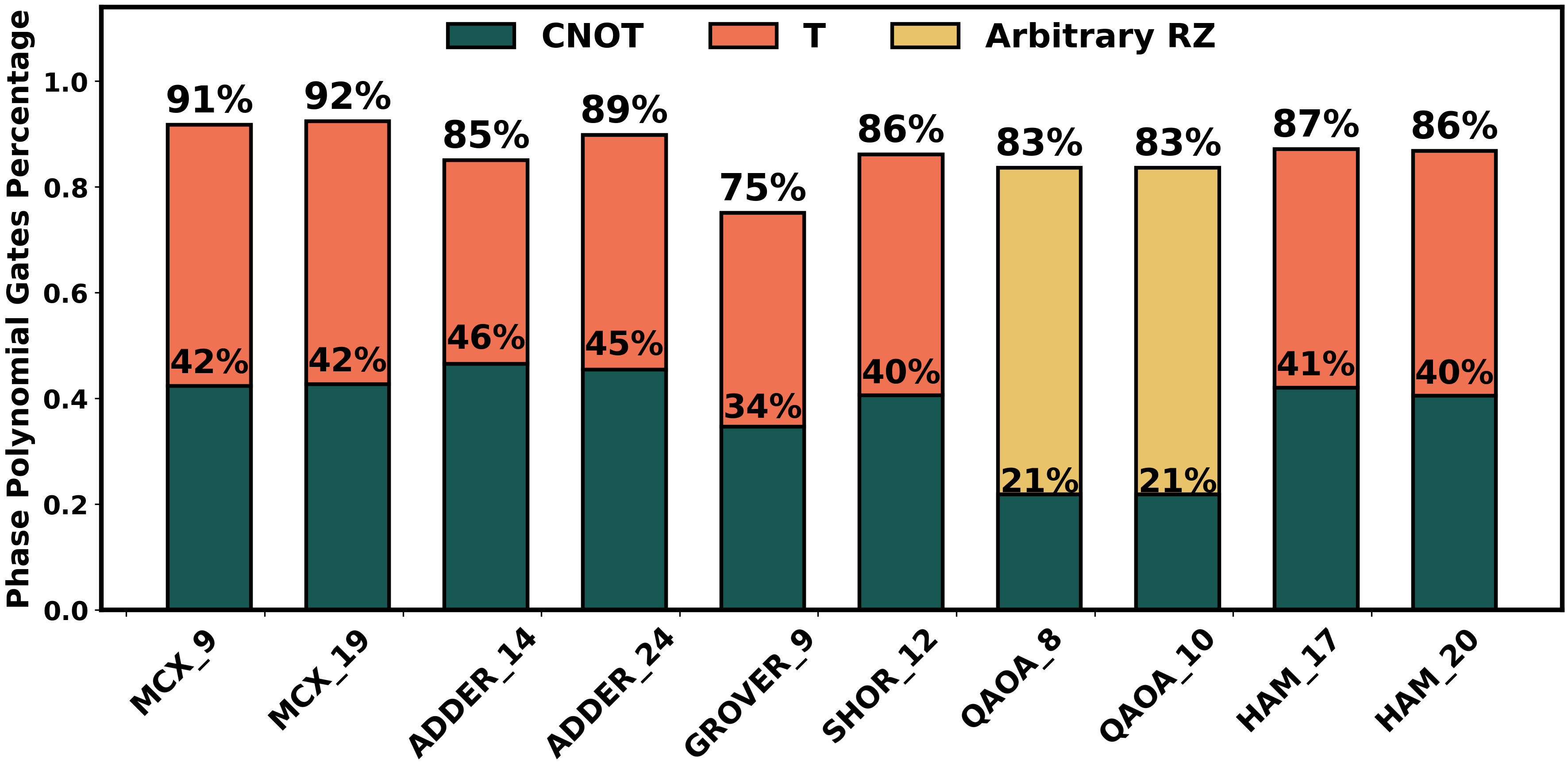}
    \caption{
Breakdown of $\{\mathrm{CNOT}, R_z\}$ usage in representative quantum circuits, showing that a large fraction of gates fall into $\{\mathrm{CNOT}, R_z\}$ regions corresponding to phase polynomial subcircuits.
The $R_z$ terms are further decomposed into $\{Z, S, T\}$, as $R_z$ subsumes both Clifford and non-Clifford rotations.
The two QAOA benchmarks represent noisy-circuit instances and therefore have arbitrary $R_z$ decompositions.
}
    \label{fig:percentage}
    
\end{figure}

\textbf{Why Are Phase Polynomials Important?}
Phase polynomial subcircuits occur extensively in quantum circuits. To quantify this, we evaluate a representative set of benchmarks commonly used in circuit optimization studies~\cite{amy2018controlled, nam2018npj, xu2022PLDI, xu2023PLDI, ruiz2025quantum}, including multi-controlled NOT (MCX) circuits~\cite{barenco1995elementary, iten2016quantum}, Grover's search~\cite{grover1996quantum}, Shor's factoring algorithm~\cite{shor1999polynomial}, the quantum approximate optimization algorithm (QAOA~\cite{jang2024recompiling}), and Hamiltonian dynamics (HAM). As shown in Fig.~\ref{fig:percentage}, more than 75\% of the gates in these benchmarks belong to $\{\mathrm{CNOT}, R_z\}$ regions, with several cases exceeding 90\%. 

\textbf{Relationship to Fault-Tolerant Quantum Computing:}
In fault-tolerant (FT) quantum computing, a standard universal gate set is Clifford+$T$, where Clifford operations are typically generated by $\{\mathrm{CNOT}, H, S\}$. Within this setting, phase polynomial regions naturally arise from the combination of CNOT gates and diagonal phase rotations (e.g., $Z$, $S$, and $T$), which together capture both parity computation and phase accumulation. As shown in Fig.~\ref{fig:percentage}, these regions dominate a wide range of FT-oriented benchmarks, where CNOT and $T$ gates account for the majority of operations and appear in comparable proportions.

Previously, FT optimization has focused primarily on reducing $T$ gates due to their high cost under magic-state distillation~\cite{bravyi2005universal}. However, recent advances in magic-state cultivation~\cite{gidney2024magic,chen2025efficient, rosenfeld2025magic} and updated resource models~\cite{huggins2025fluid} indicate that the costs of $T$ gates and CNOTs are becoming increasingly comparable. This shift highlights the need for optimization techniques that jointly reduce both gate types. In particular, it motivates treating phase polynomial regions as first-class optimization targets, as they directly capture the dominant cost structure in both FT and near-term noisy circuits.

\textbf{How Are Phase Polynomials Integrated into General Circuit Optimization Frameworks?}
Current approaches use phase polynomials as auxiliary tools for local circuit-rewriting optimizers. For instance, Quartz~\cite{xu2022PLDI} uses rotation merging (a subset of phase polynomial techniques) as a preprocessing step. Quartz automatically searches and constructs equivalent circuit classes (ECCs) for local subcircuit rewriting---replacing a sub-circuit with a better one in its ECC class. QUESO~\cite{xu2023PLDI} is also a rewriting framework, which uses phase polynomials to enhance ECC generation via a polynomial identity filter (PIF)---creating more circuits in ECCs for rewriting purposes. 

Phase polynomials are used as auxiliary tools, rather than a standalone optimization pass. This is due to the lack of a unified intermediate representation that addresses not only a single block of CNOT + $R_z$ gates, but also transitions between blocks and to other non-phase-polynomial gates. Amy \etal~\cite{amy2018controlled} propose the Gray-Synth algorithm, which greedily orders and synthesizes phase terms according to the Gray code. However, its theory applies only to single blocks and only to their phase-rotation components. It does not systematically handle how XOR propagation of phase terms interacts with the block's output-basis transformation—a crucial requirement for stitching phase polynomials correctly into a full circuit. Because Gray-Synth is restricted to individual phase polynomial blocks, it cannot operate on general quantum circuits with non-phase-polynomial gates, limiting its applicability as a general-purpose circuit optimization pass.

\textbf{How Our Work Differs from Prior Work:}
We present \emph{PhasePoly}, a compiler optimization pass that elevates phase polynomial optimization from an auxiliary technique to a first-class stage in general circuit compilation. The key insight is that phase polynomial structure enables global reasoning beyond local rewriting, if both phase and parity transformations are modeled in a unified and extensible way.

\textbf{
\emph{1. Beyond the phase-parity: A Unified Representation for Comprehensive Phase Block Analysis.}
}
Within a single $\{\mathrm{CNOT}, R_z\}$ block, prior methods primarily minimize the cost of the phase-parity network while treating the output basis transformation as a separate synthesis problem. This separation misses co-optimization opportunities: different realizations of the same phase terms can induce different downstream costs. We address this by introducing a unified representation that captures both phase and output parities under the same CNOT transformations, enabling coordinated optimization.

\textbf{
\emph{2. Breaking the Single-Block Barrier: Cross-Block Optimization.}
}
Phase polynomial regions in general circuits are partitioned by basis-changing gates (e.g., $H$), limiting existing approaches to short-range, block-local improvements. We overcome this limitation by introducing a cross-block intermediate representation and optimization that enables long-range transformations beyond ${\text{CNOT}, R_z}$ subcircuits.
As illustrated in Fig.~\ref{fig:Multi-block}, $H$ gates partition the circuit into three phase polynomial blocks. Prior approaches, such as Gray-Synth, are limited to block-local optimization, whereas our cross-block approach jointly optimizes non-adjacent blocks (e.g., the first and third), uncovering optimization opportunities beyond block boundaries while preserving correctness and efficiency.

\textbf{\emph{3. Standing Alone, Working Together: Orthogonal Integration with Other Frameworks.}} Across diverse benchmarks—including arithmetic circuits, multi-controlled Toffoli gates, Hamming coding functions~\cite{saeedi2010reversible}, Hamiltonian simulation, QAOA, Grover’s algorithm, and Shor’s algorithm— we show strong standalone performance of \emph{PhasePoly}. We also demonstrate complementary benefits when combining \emph{PhasePoly} with existing rewriting frameworks. Our results demonstrate that phase polynomial optimization should be treated as a first-class stage in the compilation pipeline: not merely an auxiliary to subcircuit rewriting frameworks, but a collaborator that exposes otherwise unreachable opportunities.

\textbf{Our contributions are summarized as follows:}

\begin{itemize}
    \item \emph{Systematic revisiting of phase polynomials:} We provide the first systematic investigation of phase polynomials in general circuit optimization, establishing their necessity as a standalone optimization pass.

    \item \emph{Holistic phase polynomial optimization:} We introduce a framework that jointly optimizes the phase rotation and output basis transformation. Combined with cross-block optimization, it overcomes single-block limitations and enables substantially stronger results.

    \item \emph{Extensibility and scalability:} Unlike fixed-size subcircuit rewriting methods, our approach scales naturally to large circuits and demonstrates strong extensibility across diverse benchmarks. Our approach delivers significant reductions in total gate count (up to 50.00\%, average 34.70\%), CNOT gates (up to 48.57\%, average 26.83\%).
    
    \item \emph{Orthogonality:} \emph{PhasePoly} is orthogonal to {subcircuit rewriting}. While rewriting may perform comparably or better on small circuits, our approach scales more effectively and uncovers additional improvement (up to 13\% for already highly optimized circuits). Together, they close both short- and long-range optimization gaps.
    
\end{itemize}

\section{Background and Key Insights}
\label{sec:mot}

\begin{figure}[htbp]
\vspace{-1\baselineskip}
\centering\includegraphics[width=0.31\textwidth]{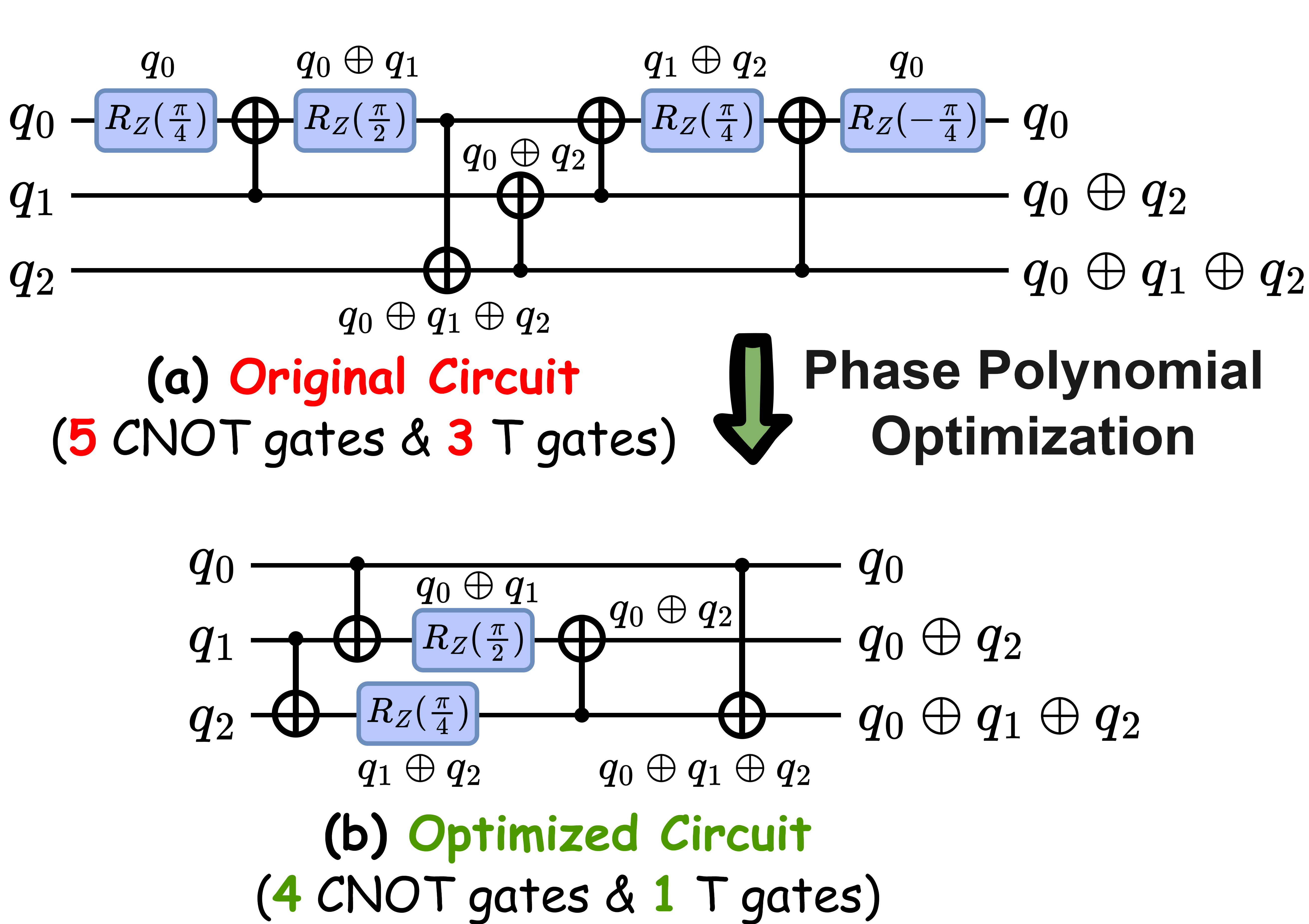}
    \caption{Phase polynomial optimization example: (a) and (b) are functionally equivalent; however, (a) uses 5 CNOTs and 3 $T$ gates, whereas (b) uses 4 CNOTs and 1 $T$ gate.}
    \vspace{-0.5\baselineskip}
 \label{fig:ppfirstexample}
\end{figure}

A quantum circuit is a sequence of quantum gates acting on an $n$-qubit system. The computational basis states are written as $\ket{x}$ with $x \in \mathbb{F}_2^n$, which is a binary vector. A CNOT acting on control $x$ and target $y$ maps $\ket{x,y}$ to $\ket{x, x \oplus y}$. An example in Fig.~\ref{fig:ppfirstexample} illustrates how CNOTs update parities and how $R_z$ gates must track the XOR sums to preserve the correct output basis. The transformed circuit is functionally equivalent to the original while eliminating 1 CNOT and 2 $T$ gates.

A \emph{phase-polynomial circuit} is a circuit region composed solely of 
$\{\text{CNOT}, R_z\}$ gates. Such regions are not universal for general quantum 
circuits. In a general circuit, the appearance of non-phase-polynomial gates 
(e.g., $H$ gates) changes the computational basis and therefore 
terminates the region. We define a \emph{phase-polynomial block} as a maximal 
contiguous subcircuit containing only $\{\text{CNOT}, R_z\}$ gates in general circuits. Formally, one can represent a phase polynomial circuit in a \emph{sum-over-paths} form~\cite{amy2014polynomial, amy2018controlled, nam2018npj} such that
\begin{align}
U\ket{x_1,\ldots,x_n} = e^{ip(x_1,\ldots,x_n)}\ket{g(x_1,\ldots,x_n)}
\label{eq:pp}
\end{align}
where $p(x)$ is a Boolean polynomial over XOR parities with phase coefficients, and 
$g(x)$ is an affine reversible transformation implemented by a CNOT network.
\begin{equation}
p(x_1,\ldots,x_n) = \sum_{y\in\{0,1\}^n} \theta_i\,\Bigl(x_1y_1 \oplus \cdots \oplus x_ny_n\Bigr)
\end{equation}

In Fig.~\ref{fig:ppfirstexample}(a), the phase function can be written as a weighted sum of parity terms: $p(q_0, q_1, q_2) = \frac{\pi}{4}q_0 + \frac{\pi}{2} (q_0 \oplus q_1) + \frac{\pi}{4} (q_1 \oplus q_2) + \frac{-\pi}{4} q_0$ $= \frac{\pi}{2} (q_0 \oplus q_1) + \frac{\pi}{4} (q_1 \oplus q_2)$.
Each term corresponds to a phase rotation conditioned on a parity of input variables. 
In general, a \textbf{phase-parity} is the XOR of a subset of input qubits, and a 
\textbf{phase-parity function} $p(x)$ is a linear combination of such parities with rotation angles.
At the circuit level, these parities are constructed using CNOT gates and realized by applying
$R_z(\theta)$ rotations on the corresponding qubit lines; we refer to this structure as the 
\textbf{phase-parity network}.

The function $g(x)$ represents the \textbf{output basis transformation}, a linear reversible 
mapping of computational basis states implemented by a CNOT network. 
For example, in Fig.~\ref{fig:ppfirstexample}(a), $g(q_0,q_1,q_2)=(q_0,\; q_0\oplus q_2,\; q_0\oplus q_1\oplus q_2).$
Each output is a parity of the input qubits. 
We call these parities \textbf{output parities}, and the corresponding CNOT circuit implementing
this linear transformation the \textbf{output-parity network}.

\subsection{Single-block Optimization: One Stone Two Birds}

The phase function $p$—capturing phase parities—has been extensively studied in the context of phase polynomial optimization~\cite{amy2014polynomial, amy2018controlled, vandaele2022QuantumScienceTechnology}. 
In contrast, the output transformation $g$—capturing output parities—is typically studied in a different context, namely linear reversible circuit synthesis~\cite{patel2008optimal, de2021gaussian}. 
No prior work has come up with a way to unify these two problems into one model; they have addressed these two problems separately and solved them one after another. 

However, such separate handling may miss co-optimization opportunities. This is because the CNOT network synthesis for the phase parity function affects the parity state of each qubit, which is subsequently used as input to the output parity component---the $g$ function.
Thus, implementations that achieve the minimal gate count for the phase parity function may not minimize the gate count for the output parity function.

\begin{figure}[t!]
    \vspace{-1.0\baselineskip}
    \centering
\includegraphics[width=0.45\textwidth]{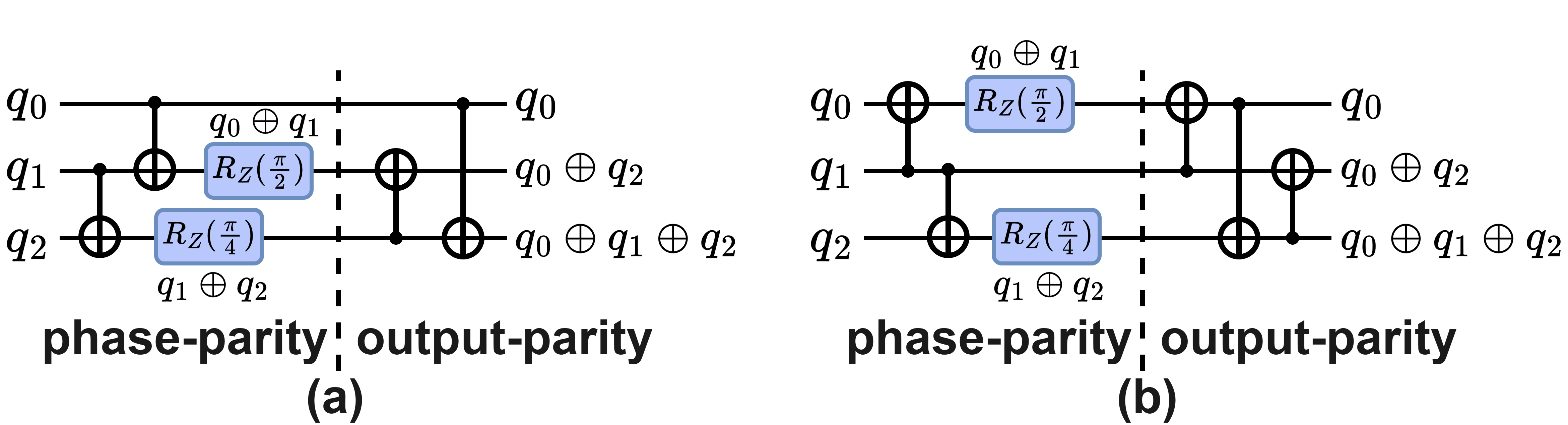}
    \caption{Two circuits that both implement the $p$ function using the same minimal gate count, but result in different costs for the $g$ function. (a) uses one fewer CNOT than (b).}
    \label{fig:Co-optimize}
     \vspace{-1\baselineskip}
\end{figure}

We show such an example in Fig.~\ref{fig:Co-optimize} where two circuits implement the phase-parity ($p$) function with the same minimal cost---both using only 2 CNOTs in (a) and (b). However, they lead to different CNOT costs in the basis transformation ($g$) function---one using two and the other using three. Thus, even if the phase-parity network is individually minimal, ignoring its interaction with the output-parity network leads to non-minimal overall CNOT overhead.
\textbf{In this paper, we unify these two problems into one, in order to capture their correlation.} For the example in Fig.~\ref{fig:Co-optimize}, our framework is able to find the overall minimal transformation cost in (a). The details of the co-optimization are in Section \ref{sec:coopoverview}. 

\begin{figure*}[t!]
    \centering
    \vspace{-1\baselineskip}
    \includegraphics[width=0.9\textwidth]{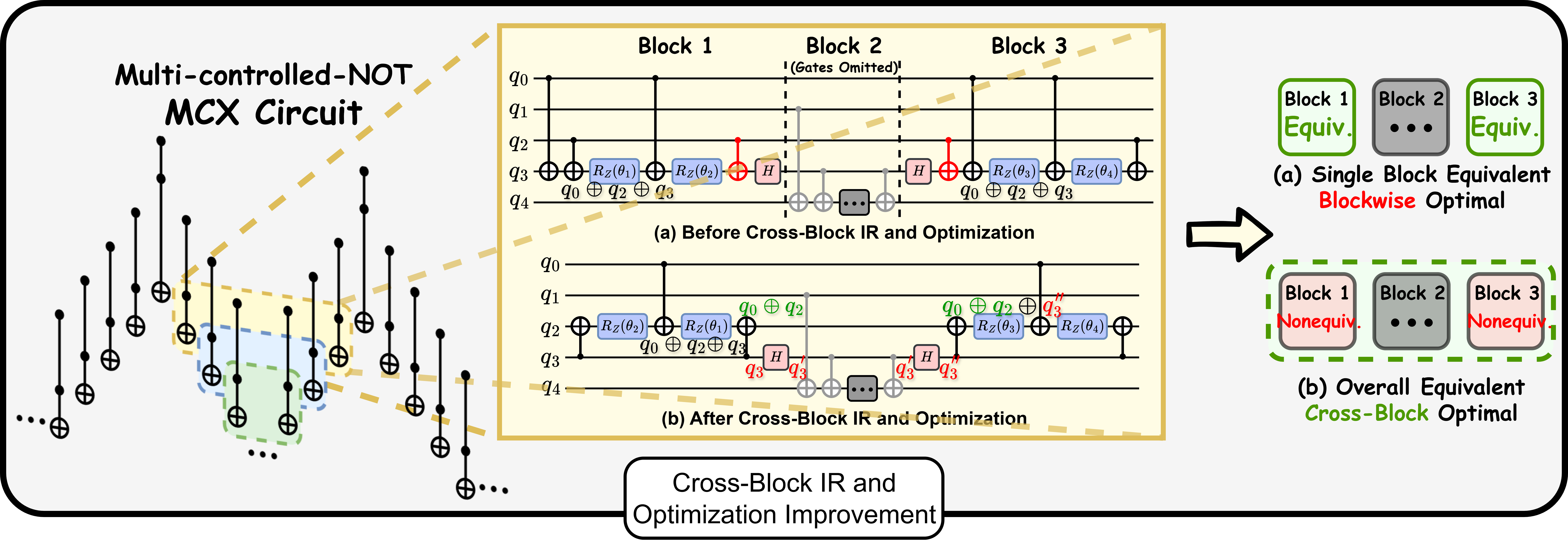}    
    \caption{
    Phase polynomial cross-block intermediate representation (IR) and optimization on a multi-controlled-NOT (MCX) circuit. 
    \emph{Left:} a portion of an MCX construction, where Toffoli gates are chained in a prescribed order over control, target, and ancilla qubits; the highlighted region (with intermediate gates omitted for clarity) is expanded on the right. 
    \emph{Circuit (a), before:} two $H$ gates act as block barriers, partitioning the circuit into three phase polynomial blocks. Optimizing each block in isolation attains block-local optima (e.g., eight CNOTs and four $R_z$ gates in the shown region) but misses cross-block reductions. 
    \emph{Circuit (b), after:} phase polynomial cross-block IR merges the three blocks into a single phase polynomial region. Cross-block optimization reorders the parity network structure and eliminates two redundant CNOT gates (marked in red in circuit (a)), preserving functional equivalence while lowering the overall CNOT cost.}
    \vspace{-1\baselineskip}
    \label{fig:Multi-block}
\end{figure*}

\subsection{Breaking the Block Barrier: Long-Range Optimizations}

A general circuit may contain non-phase polynomial gates (e.g., $H$ gates) acting as \emph{block barriers} that split a circuit into multiple phase polynomial blocks. Optimizing individual blocks may be insufficient, as each phase polynomial block may be too small. \textbf{This raises a key question: can optimization opportunities be exposed across block boundaries?}

We show that the answer is YES. Consider the multi-controlled-NOT gate (MCX)~\cite{barenco1995elementary, iten2016quantum}, a core primitive in many algorithms and simulations~\cite{zindorf2024efficient,arrazola2022universal,grover1996quantum,shor1999polynomial}, which appears prominently in modular exponentiation—the dominant cost in Shor’s algorithm. We show an MCX implementation with n qubits using the standard 3-qubit Toffoli gates in Fig.~\ref{fig:Multi-block} (left). This implementation is further decomposed into Clifford and $T$ gates. The $H$ gates act as barriers, forcing phase polynomial optimization to act only within a Toffoli gate block in the traditional phase polynomial context. 

Traditionally, both local rewriting and single-block phase-polynomial methods reduce a Toffoli block's CNOT cost from 6 to about 4 on average. Our cross-block approach goes further: by breaking the block barrier, we reduce roughly half of the Toffoli structures to 3 CNOTs on average, achieving a non-constant improvement in CNOT count. In Fig.~\ref{fig:Multi-block}, we show the circuit in the highlighted regions, for which a parity term $q_0 \oplus q_2$ can be reused (generated in block 1 and reused in block 2). This eliminates two CNOT gates across two blocks, reducing the cost by one CNOT per block on average. This opportunity, however, cannot be captured by the traditional approach, as (1) between blocks 1 and 3, there are many other gates---exceeding the local sub-circuit rewriting range, and (2) $H$ gates delineate the single block boundaries.

We further provide quantitative evidence in Fig.~\ref{fig:motivation_mcx}, showing MCX circuits of increasing size. 
Our approach achieves greater reductions in both CNOT and total gate counts compared to Quartz~\cite{xu2022PLDI} and QUESO~\cite{xu2023PLDI}, with the gap widening as the number of qubits increases. 
This trend arises because the number of Toffoli gates scales with circuit size, increasing the opportunities for cross-block parity reuse. These results show that our approach captures optimization opportunities beyond prior techniques. 
Details of the cross-block optimization are described in Section~\ref{sec:cross-block}.

\subsection{Overcoming the Scaling Challenges}
\label{sec:long-range}

State-of-the-art quantum circuit optimizers largely rely on \emph{equivalent subcircuit rewriting}, where small subcircuits are replaced using precomputed equivalence classes (ECCs). 
While effective for local transformations, these methods are fundamentally limited by the size of the rewrite rules: practical ECCs typically cover only small patterns (e.g., up to 3 qubits and depth 3--6), as constructing larger equivalence classes becomes computationally intractable. 
As a result, capturing long-range optimizations requires many local rewrites and becomes increasingly difficult as circuit size grows.

In contrast, our approach avoids this limitation by operating on a structured parity-matrix representation rather than fixed-size rewrite patterns. By manipulating this representation for both single- and cross-block optimization (see Section~\ref{sec:logopt}), it scales naturally with circuit size and captures long-range transformations beyond local rewriting.

\begin{figure}[htbp]
    \vspace{-0.5\baselineskip}
    \centering
    \includegraphics[width=0.5\linewidth]{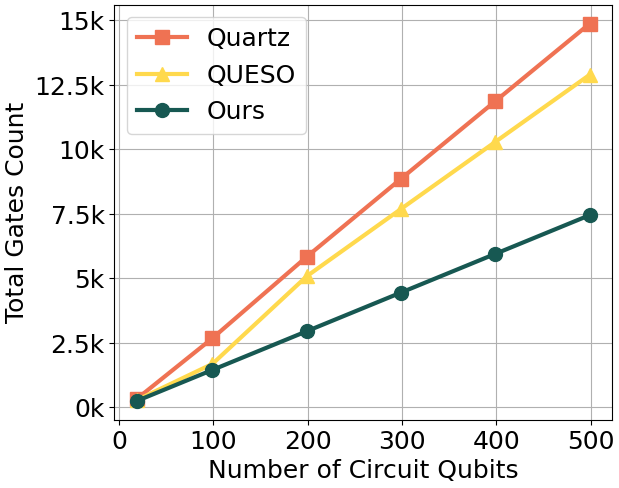}
    \hfill
    \includegraphics[width=0.47\linewidth]{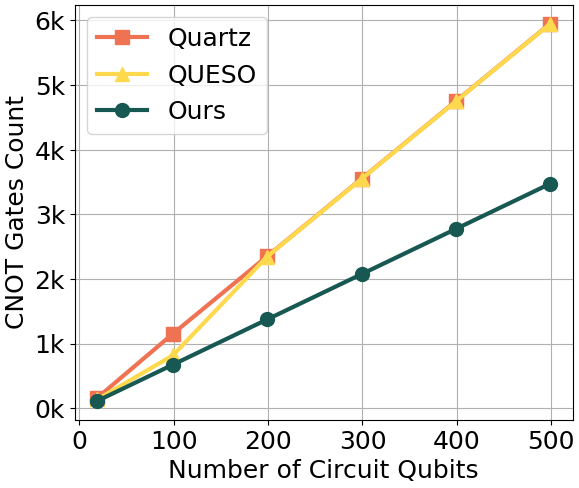}
    \caption{
        Gate count comparison of different optimization frameworks on multi-controlled NOT (MCX) circuits. 
    }
    \label{fig:motivation_mcx}
    \vspace{-0.5\baselineskip}
\end{figure}

We compare against subcircuit rewriting frameworks Quartz and QUESO on MCX circuits (Fig.~\ref{fig:Multi-block}, left). 
Rewriting methods are given a fixed 7200s budget per circuit, while our phase polynomial optimization uses at most 3600s for the largest instances. 
Within these settings, the performance of local subcircuit rewriting degrades steadily as the number of qubits grows from 19 to 100, with even sharper declines at larger scales (Fig.~\ref{fig:motivation_mcx}). Since QUESO extends subcircuit rewriting by incorporating phase contributions, its performance is slightly weaker than phase polynomial optimization but consistently stronger than Quartz for circuits with fewer than 100 qubits. However, as circuit scale increases, the gap widens: QUESO's reduction rate deteriorates, while phase polynomial optimization continues to sustain a linear growth trend.

\section{Phase Polynomial Circuit Optimization}
\label{sec:logopt}

\subsection{Co-Optimization of Phase- and Output-parity Networks}
\label{sec:coopoverview}

The phase polynomial optimization consists of two components:
the \emph{phase-parity} network and the \emph{output-parity} network.
Prior work typically optimizes them separately. The output-parity network corresponds to the linear transformation
$g(x)$ in Eq.~\ref{eq:pp}. This transformation can be represented
as a binary matrix over $\mathrm{GF}(2)$, where each row encodes
the parity appearing in the corresponding output qubit.

Each CNOT corresponds to an elementary row operation over $\mathrm{GF}(2)$:
a $\mathrm{CNOT}(i,j)$ performs
$\text{row}_j \leftarrow \text{row}_j \oplus \text{row}_i$.
Thus, a CNOT circuit can be viewed as a sequence of matrix updates that,
starting from the identity, produce the output-parity matrix describing
$g(x)$.

\begin{figure}[h]
    \vspace{-1\baselineskip}
    \centering
\includegraphics[width=0.5\textwidth]{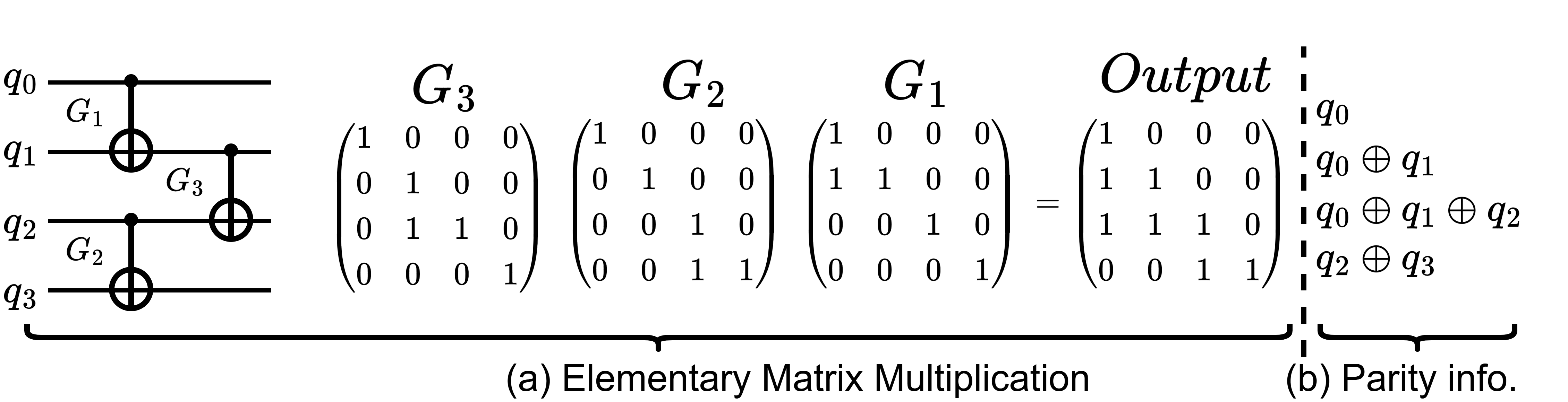}
\caption{
CNOT gates correspond to elementary row operations over $\mathrm{GF}(2)$.
Synthesizing the CNOT network is equivalent to reducing this matrix to the identity via Gaussian elimination.
}
\label{fig:glreasoning}
    \vspace{-0.5\baselineskip}
\end{figure}

For example, in Fig.~\ref{fig:glreasoning}, $G_1 \mapsto \mathrm{CNOT}(q_0,q_1)$
updates the row of $q_1$ to $q_0\oplus q_1$, represented as $[1,1,0,0]$ in the second row of the matrix $G1$.
Conversely, synthesizing a CNOT network for a given $g(x)$ amounts to
reducing the matrix back to the identity via elementary matrix multiplications,
which can be performed using Gaussian elimination or related linear-algebra techniques~\cite{patel2008optimal, de2021gaussian}, which guarantees that the transformation can be realized in polynomial time.

However, this representation cannot directly capture phase-parity networks with parity-controlled $R_z$ rotations. 
Unlike the output-parity network, the phase-parity block is not a square transformation matrix: each column represents a parity term rather than an output mapping, and therefore it cannot be reduced to the identity using Gaussian elimination.

To jointly optimize the phase-parity and output-parity networks, we represent both in a unified column-based parity form. 
Each phase term corresponds to a parity vector (e.g., $(110)^T$ denotes $q_0 \oplus q_1$). 
The output transformation $g(x)$, typically represented as a square matrix (e.g., Fig.~\ref{fig:glreasoning}), can be transposed so that each output parity is also a column vector. 
Thus, both $p(x)$ and $g(x)$ are expressed as collections of parity vectors over the input qubits.

We therefore introduce a \emph{coupled parity matrix} representation:
$[\;\text{phase-parity block} \mid \text{output-parity block}\;]$, to jointly represent
phase terms and output parities. Under our convention, $\mathrm{CNOT}(i,j)$ applies
$\mathrm{row}_i \leftarrow \mathrm{row}_i \oplus \mathrm{row}_j$ to both
blocks simultaneously, coupling their optimizations. Though a physical CNOT updates the target qubit, our operation updates the phase-parity term instead of the quantum state itself, following prior works~\cite{amy2018controlled,vandaele2022QuantumScienceTechnology}.

Fig.~\ref{fig:cnotyx} illustrates this representation for the circuit in Fig.~\ref{fig:ppfirstexample}. 
When a phase-parity column reaches Hamming weight~1, its parity depends on a single qubit, so the corresponding $R_z$ rotation can be emitted and the column removed. 
For example, after $\mathrm{CNOT}(q_1,q_0)$, the column $(110)^T$ becomes $(100)^T$, meaning that the parity $q_0 \oplus q_1$ has been propagated onto qubit~0. Repeating such row operations eliminates phase columns while updating the output-parity matrix.
Once all phase columns are removed, the remaining output-parity matrix can be synthesized via Gaussian elimination.

\begin{figure}[htbp]
        \vspace{-0.5\baselineskip}

    \resizebox{1.0\linewidth}{!}{
        $
        \begin{bmatrix}[cc|ccc]
        1 & 0 & 1 & 1 & 1 \\
        1 & 1 & 0 & 0 & 1 \\
        0 & 1 & 0 & 1 & 1 \\
        \end{bmatrix}
        \xrightarrow{\text{CNOT}(q_{1}, q_{0})}
        \begin{bmatrix}[cc|ccc]
        1 & 0 & 1 & 1 & 1 \\
        0 & 1 & 1 & 1 & 0 \\
        0 & 1 & 0 & 1 & 1 \\
        \end{bmatrix}
        \xrightarrow{\text{Insert } R_z}
        \begin{bmatrix}[c|ccc]
        0 & 1 & 1 & 1 \\
        1 & 1 & 1 & 0 \\
        1 & 0 & 1 & 1 \\
        \end{bmatrix}
        $
    }
    \caption{
        Coupled co-optimization representation. 
        A CNOT simultaneously updates the phase-parity and output-parity block. 
    }
        \vspace{-0.5\baselineskip}
    \label{fig:cnotyx}
\end{figure}

\subsection{Overall Phase Polynomial Co-Optimization Framework}

This optimization task can be formulated as a \textbf{CNOT-minimization parity network synthesis problem}~\cite{amy2018controlled}, where the optimization objective is to minimize the total gate count.
CNOT network synthesis is NP-hard~\cite{amy2018controlled, van2023optimising} for both phase-parity and output-parity optimization, making it infeasible to guarantee optimal solutions in polynomial time. We therefore adopt a heuristic-based framework to optimize the sequence of CNOT and $R_z$ operations.

We model the search space as a tree in which each node represents a circuit state and each edge corresponds to a CNOT operation (Fig.~\ref{fig:treeStructure}). A root-to-leaf path specifies a sequence of CNOTs. A valid leaf node is reached when the phase-parity matrix becomes empty, and the output-parity matrix reduces to the identity. The optimization objective is to minimize the number of CNOT gates, which corresponds to finding a minimum-cost path in this search tree.

\subsubsection{Active Row Pair CNOT Selection}
\label{sec:activeRowPair}

As described in Section~\ref{sec:coopoverview}, a CNOT corresponds to an XOR between two rows of the parity matrix, referred to as a \textbf{row pair}. The operation is directional: a pair $(\text{row}_i,\text{row}_j)$ represents a CNOT with control qubit $i$ and target qubit $j$, updating the control row as $\text{row}_i \oplus \text{row}_j$ while leaving the target row unchanged.

A row pair is considered valid if it reduces the Hamming weight of at least one column. Such pairs are called \textbf{active row pairs}. However, repeatedly applying the same row pair can cause livelock, preventing progress. To avoid this, we restrict the search using an \textbf{active column set}, defined as the subset of columns whose Hamming weights can be reduced. Row pairs are only considered if they help at least one column within the active column set (reducing at least one 1). After a CNOT operation, the active column set may be updated. At initialization, or after eliminating a column, the active column set includes all remaining columns.

\subsubsection{The Priority Queue Implementation}
\label{sec:priorityQueue}

To evaluate the state of each move after applying a CNOT from the active row pair set, we derive the following insights:

\begin{itemize}
    \item \textbf{Phase-parity cost $h_1(n)$:} the total Hamming weight (number of 1s) in the phase-parity matrix correlates with the number of CNOT gates required for phase terms.
    \item \textbf{Output-parity cost $h_2(n)$:} the estimated number of CNOTs required to transform the output-parity matrix into the identity using Gaussian elimination.
    \item \textbf{Accumulated cost $g(n)$:} the number of CNOTs already applied along the path from the root to the current node.
\end{itemize}

The overall cost function is:
\begin{equation}
f(n) = g(n) + h_1(n) + h_2(n)
\end{equation}

\begin{figure}[t!]
    \vspace{-1.0\baselineskip}
    \centering
    \includegraphics[width=0.48\textwidth]{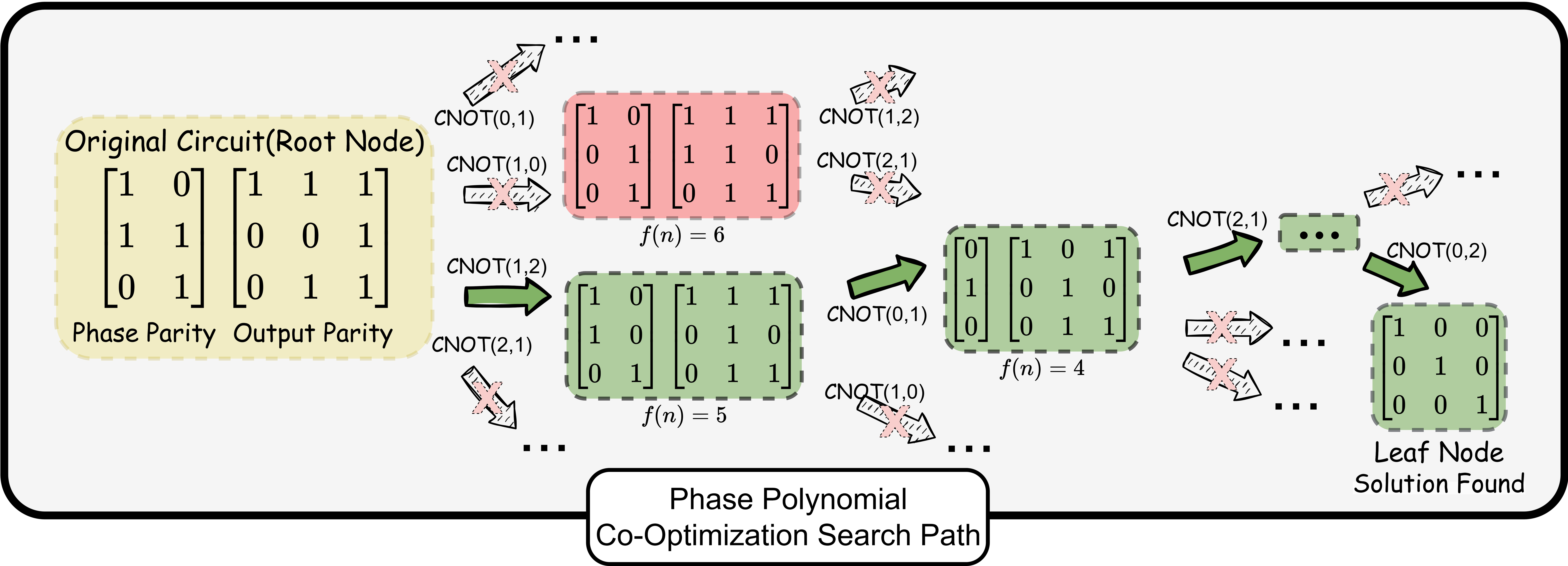}
    \caption{
    Search tree for Phase Polynomial Co-Optimization.  
    The root denotes the original circuit, the leaf represents the optimized circuit, and edges represent CNOT operations.  
    Node costs reflect heuristic evaluations of intermediate states.}
    \label{fig:treeStructure}
    \vspace{-1\baselineskip}
\end{figure}

We adopt an A* search with a priority queue ordered by $f(n)$. When a state reaches an empty phase-parity matrix, the synthesis of the output-parity network is completed using Gaussian elimination. Otherwise, new successor states are generated by applying CNOTs from the active row pair set in an effort to empty the phase-parity matrix. An example is illustrated in Fig.~\ref{fig:treeStructure}.

When multiple states share the same cost, we apply a tie-breaking rule based on the lexicographic ordering of
\[
[f(n),\; h_1(n),\; h_2(n),\; -g(n)].
\]
This rule prioritizes states with lower estimated costs and, secondarily, those closer to completion.

Because the search space grows exponentially, unrestricted A* search becomes impractical. We therefore use a \textbf{space-bounded A* search}~\cite{russell1992efficient,kaindl1994AAAI}, capping the priority queue size. When the limit is reached, lower-priority nodes are discarded to prune unpromising paths and control memory growth. We further employ a \textbf{multiple-solution search} controlled by a hyperparameter $k$. Each time a goal state is found, it is added to a solution set of size $k$. The search terminates when either the priority queue is empty or $k$ solutions are obtained, after which the best one is returned.

\subsection{Cross-block Intermediate Representation and Optimization}
\label{sec:cross-block}

\subsubsection{Static Single-assignment (SSA) Style Rotation Merging}
\label{sec:ssa_rm}

\begin{figure}[t!]
    \vspace{-1.0\baselineskip}
    \centering
    \includegraphics[width=0.49\textwidth]{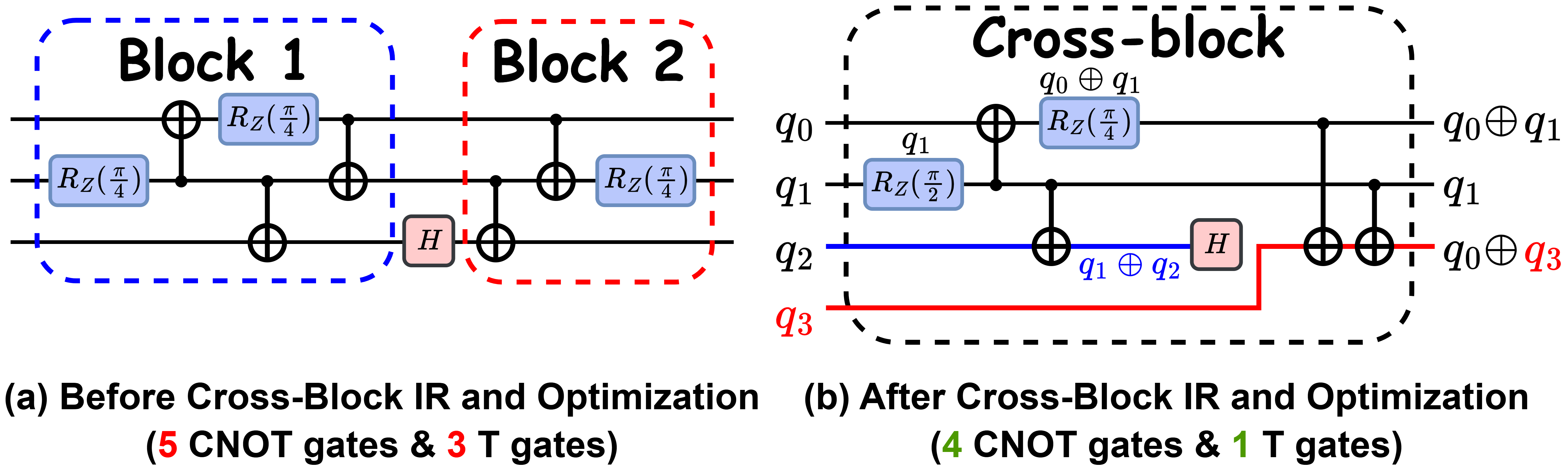}
    \caption{
        Cross-block  optimization:
        (a) Two separate blocks before optimization (5 CNOTs, 3 $T$ gates).
        (b) Across the $H$ gate, we create a new qubit wire. In the new circuit, the output-parity before the $H$ gate is set to $q_1\oplus q_2$, the same as that before the $H$ gate in the original circuit. Now we have a new phase-polynomial block, by optimizing this new block, we reduce the circuit to 4 CNOTs and 1 $T$.
    }
        \vspace{-1\baselineskip}
\label{fig:cross_block_tech}
\end{figure}

For the standard Clifford+$T$ gate set $\{T,T^{\dagger},S,S^{\dagger},H,X,\mathrm{CNOT}\}$~\cite
{gottesman1998theory}, the phase polynomial gate set $\{\mathrm{CNOT},R_z\}$ is \emph{not} universal. We therefore partition a general quantum circuit into blocks of phase polynomial subcircuits separated by non-$R_z$ single-qubit gates. In particular, $H$ gates act as \emph{block barriers}: two-qubit gates whose semantics depend on the post-$H$ basis are excluded from the preceding phase polynomial block. Fig.~\ref{fig:cross_block_tech}(a) illustrates a case with two such blocks.

To enable rotation merging across block boundaries, we introduce an SSA-style \cite{rastello2022ssa} \emph{qubit-state renaming and rotation merging}.  
Each input qubit state, and each state created after an $H$ gate, is assigned a fresh SSA identifier. 
Every $R_z$ gate is then tagged with the SSA ID of the qubit state it acts on. 
By merging all phase terms associated with the same SSA ID, we achieve \emph{whole-circuit} rotation merging rather than being restricted to a single block.

For example, in Fig.~\ref{fig:cross_block_tech}(b), two $T$ gates on $q_1$ share the same SSA ID and therefore merge, even though they appear in different blocks. 
Likewise, the blue wire $q_2$ ends at the SSA state $\{q_1\oplus q_2\}$ before the $H$ gate;
after the $H$ gate we create a new SSA state~$q_3$, yielding an updated output-parity $\{q_0 \oplus q_3\}$.

Prior work on rotation merging~\cite{nam2018npj} typically operates within a single block, using \emph{anchors} and \emph{terminal points} to extend block-local cancellations; this increases implementation complexity and provides limited guarantees~\cite{hietala2021POPL}. In contrast, SSA-based qubit-state renaming yields a simple correctness criterion—rotations merge \emph{iff} they target the same SSA ID—thereby enabling global merging across block boundaries. 
Recent work by Amy and Lunderville~\cite{amy2025linear} formulates rotation merging via relational program analysis, discovering additional opportunities across control flow and non-linear relations (Toffoli gates) for rotation gate merging. 
Our approach is complementary: it also exposes parity relationships between two-qubit gates, constructing a \textbf{Cross-block IR} that reveals long-range parity reuse and reduces CNOTs across blocks.

Before constructing rotation merging, we perform preprocessing to reduce redundant block barriers:
(i) Propagate $X$ gates forward via Clifford conjugation so that $H$ gates remain the only block barriers;
(ii) Cancel adjacent $H$ pairs;
(iii) If a CNOT gate is bracketed by $H$ on both wires, cancel the four $H$ gates and switch the control-target accordingly; if exactly one wire is bracketed, conjugate that wire and insert two $H$ gates on the other wire to preserve equivalence.
We interleave this preprocessing pass with rotation merging twice.

\subsubsection{Cross-block Parity Matrix Intermediate Representation Design}
\label{sec:cross-block-ir}We merge adjacent single-block phase polynomial regions into a larger phase polynomial block using a \emph{cross-block IR}, as shown in Fig.~\ref{fig:cross_block_tech}(b). 
In the cross-block setting, post-$H$ qubit states (new qubit rows) exist in the IR but remain \emph{inactive} until their producer row (the original qubit wire before the H) is eliminated and the $H$ gate is inserted; only then is the row activated and available for operations.

To eliminate a pre-$H$ row  and activate its successor (e.g., $q_3$ in Fig.~\ref{fig:cross_block_tech}), three conditions must hold:
\begin{enumerate}
  \item \textbf{No pending phase terms on the row:} the corresponding row in the phase-parity block is all zeros (no remaining phase terms depend on this state).
\item \textbf{Column isolation:} in the output-parity block, if the $i$-th row is to be removed, the $i$-th column must form a unit vector with its sole $1$ located at row $i$ (indicating that the target output state has been correctly prepared).

\item \textbf{Row isolation:} in the output-parity block, the $i$-th row must be a unit vector, containing a single $1$ at its diagonal position in the output matrix.
\end{enumerate}

When these are satisfied, this pre-$H$ row can be removed, which activates the post-$H$ row. The pre-$H$ row maintains the correct output-parity, i.e., $q_1 \oplus q_2$ in Fig.~\ref{fig:cross_block_tech}(b), before it retires. 
Elimination of the pre-$H$ row does not cause information loss: any correlations between the eliminated row and others have already been addressed or transferred. 
All subsequent transformations remain row operations induced by CNOTs.

\subsubsection{Linear Dependency Check for Correctness}
\label{sec:cross-block-ir-linear}
Once conditions (1) and (3) are satisfied, condition (2) follows directly. 
In such a case, the joint parity matrix contains a row of the form \(v=[\,0\cdots0 \mid 0\cdots010\cdots0\,]\). 
Because all other entries of \(v\) are zero, adding this row to any other row clears the \(1\)'s in the same column, achieving column isolation. 
Therefore, we first make sure that the CNOT network synthesis process can produce circuit states that satisfy condition (3) together with condition (1), producing the required diagonal \(1\) in the output block, which is then used to enforce condition (2). 
Making sure these two conditions are satisfiable, therefore, is equivalent to checking whether the target row $v$ can be written as an XOR (over $\mathrm{GF}(2)$) of a set of candidate rows from the overall parity matrix. 
Equivalently, we must check whether the desired unit vector~\(v\) lies in the span of these rows.

We use a rank-based test for this purpose.
Let \(M\) be the matrix formed by the candidate rows. 
Appending \(v\) as an additional row, elimination of the pre-$H$ row is
feasible \emph{iff} $\mathrm{rank}(M \cup \{v\}) = \mathrm{rank}(M)$,
otherwise \(v\) is linearly independent of \(M\) and the elimination is
impossible.

Hence, as we update the coupled matrix \([\,\text{phase-parity}\mid\text{output-parity}\,]\) using CNOT gates, we constantly check whether the resulting parity matrix passes the rank test. 
If a state fails this rank test, it is immediately pruned. 
This check ensures that the pre-$H$ row(s) are removable in the end, and significantly improves efficiency.

\subsubsection{Cross-Block Optimization Implementation}
Using the above representation and rank check, we integrate it with the co-optimization of the phase-parity network and the output-parity network (Section~\ref{sec:coopoverview}). 
Post-$H$ SSA rows remain \emph{inactive} and are excluded from row-pair selection until their pre-$H$ producer row satisfies conditions (1), (2), and (3).
Then, we insert the $H$ gate, and \emph{activate} its post-$H$ row.

\paragraph{Feasibility and witness set}
We first run the linear dependency check (Section~\ref{sec:cross-block-ir-linear}); if it is infeasible, we prune the state. 
When feasible, solve $M\alpha=v$ over $\mathrm{GF}(2)$ and let $S=\{i\mid \alpha_i=1\}$ be a \emph{witness set} whose XOR equals $v$. 
Write $t$ for the index of the pre-$H$ row to eliminate. 
If $t\notin S$, select any $j\in S$ and apply $\mathrm{CNOT}(j,t)$. 
This replaces $\text{row}_j$ by $\text{row}_j\oplus \text{row}_t$ without changing the span, yielding an equivalent witness set $S'$ with $t\in S'$.
\emph{Row isolation (condition~(3)).} 
For each $i\in S'\setminus\{t\}$, apply $\mathrm{CNOT}(t,i)$ so that $\text{row}_t\leftarrow \bigoplus_{i\in S'} \text{row}_i = v$.
\emph{Column isolation (condition~(2)).} 
For every $k\neq t$ with a $1$ in column $t$, apply $\mathrm{CNOT}(k,t)$ to clear that entry.
\emph{Eliminate and activate.} 
Remove row $t$ and insert $H$, which activates the post-$H$ SSA row. 
Fig.~\ref{fig:cross_block_tech_example_2} illustrates a case with $t\notin S$: we first bring $t$ into the witness with one CNOT, then isolate the row, clear the column, and finally eliminate $t$ and insert $H$ to activate the post-$H$ state.

\begin{figure}[htbp]
    \vspace{-0.5\baselineskip}
    \centering
    \resizebox{\linewidth}{!}{%
      \(
        \left[
        \begin{array}{cccc}
          0 & 1 & 0 & 0 \\
          1 & 0 & 1 & 0 \\
          0 & 0 & 1 & 0 \\
          0 & 0 & 0 & 1 \\
        \end{array}
        \right]
        \xrightarrow{\mathrm{CNOT}(q_{1},q_{0})}
        \left[
        \begin{array}{cccc}
          0 & 1 & 0 & 0 \\
          1 & 1 & 1 & 0 \\
          0 & 0 & 1 & 0 \\
          0 & 0 & 0 & 1 \\
        \end{array}
        \right]
        \xrightarrow{\substack{\mathrm{CNOT}(q_{0},q_{1}) \\ \mathrm{CNOT}(q_{0},q_{2}) \\ \mathrm{CNOT}(q_{1},q_{0})}}
        \left[
        \begin{array}{cccc}
            \textbf{1} & \textbf{0} & \textbf{0} & \textbf{0} \\
            \textbf{0} & 1 & 1 & 0 \\
            \textbf{0} & 0 & 1 & 0 \\
            \textbf{0} & 0 & 0 & 1 \\
        \end{array}
        \right]
        \xrightarrow{\substack{\text{Eliminate } q_{0} \\ \text{Insert } H}}
        \left[
        \begin{array}{ccc}
            1 & 1 & 0 \\
            0 & 1 & 0 \\
            0 & 0 & 1 \\
        \end{array}
        \right]
      \)%
    }
    \caption{
    Example of the $t\notin S$ case, $t=1$ is the row to be eliminated, and $S$ is the witness set which will XOR to the target row.
    \textbf{(A)} Initial cross-block intermediate representation, currently, the first row is expected to be eliminated, and the fourth row will be activated after that.
    \textbf{(B)} Bring $t$ into the witness combination, yielding an equivalent witness set $S'$ with $t\in S'$.
    \textbf{(C)} Isolate row/column $t$ using CNOTs. 
    \textbf{(D)} Eliminate $t$ and insert $H$ to activate the post-$H$ state.
    }
    \label{fig:cross_block_tech_example_2}
    \vspace{-0.5\baselineskip}
\end{figure}

\paragraph{Scalability and robustness}  
The cross-block mechanism ensures correctness by locking inactive post-$H$ rows and pruning infeasible states, but merging blocks increases complexity and may occasionally underperform the single-block optimization. 
To address this, we apply two strategies. 

First, block merging is performed during a forward traversal of the block DAG obtained after applying the partition rules in Section~\ref{sec:ssa_rm}. 
Each block corresponds to a node, with edges representing qubit dependencies. 
At each node, we evaluate merge opportunities with its immediate predecessor blocks.

To control complexity, we apply pruning: two blocks are merged only if they share multi-qubit interactions. 
The merging terminates when no further sharable interactions exist or when the merged block reaches the predefined size limit. 
For example, in Fig.~\ref{fig:cross_block_tech}, Block~1 acts on $\{q_0,q_1,q_2\}$ and Block~2 on $\{q_0,q_1,q_3\}$. Since they share $\{q_0,q_1\}$, merging is allowed.

Second, we adopt an \emph{Incremental Block Merging} strategy: we begin with single-block optimization as a stable baseline, then progressively apply cross-block merging
gradually and interleave with local refinement. 
If a merge fails to improve upon the previous, we revert it and keep the previous state.

    \begin{figure*}[t!]
            \vspace{-1\baselineskip}
        \centering    
        \includegraphics[width=1.0\textwidth]{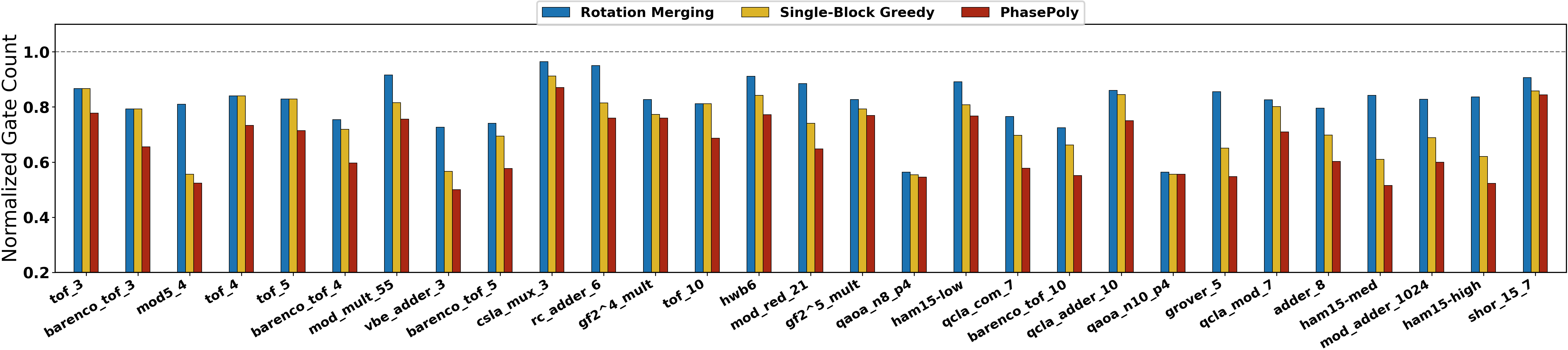}
        \vspace{-1\baselineskip}
    
        \includegraphics[width=1.0\textwidth]{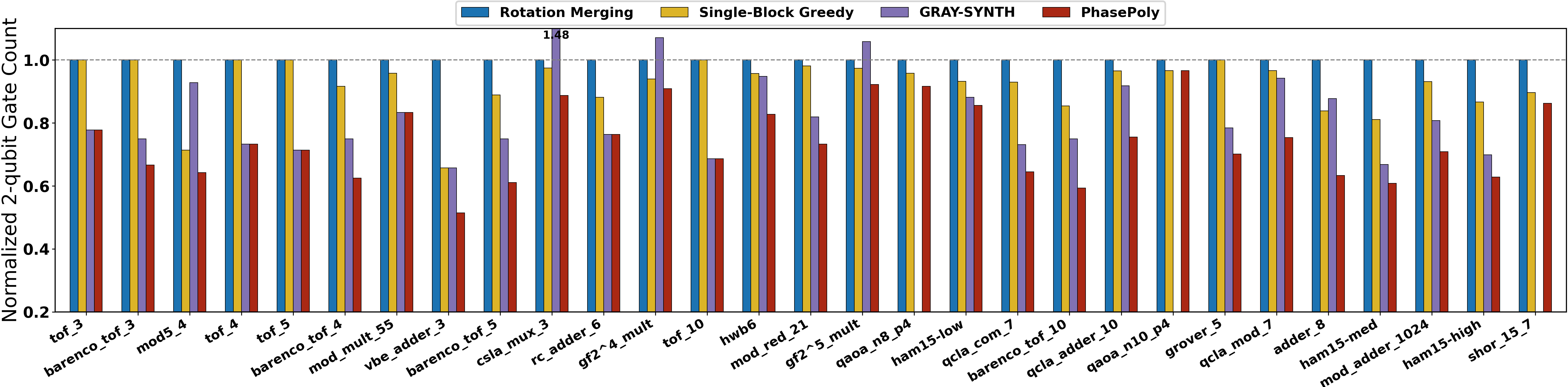}
        \vspace{-1.5\baselineskip}
        \caption{
        Normalized total and two-qubit gate-count reductions across benchmark circuits, comparing \emph{PhasePoly} against phase polynomial baselines. All values are normalized to the unoptimized circuits (1.0), with lower bars indicating greater reduction. 
        }
        \vspace{-0.5\baselineskip}
        \label{fig:table1}
    \end{figure*}

    \section{Evaluation}

We implement the proposed techniques in \textbf{\emph{PhasePoly}} and evaluate them through the following research questions:

\noindent\textbf{Q1: How does \emph{PhasePoly} compare with existing phase polynomial optimization methods?}

\noindent\textbf{Q2: Why is \emph{PhasePoly} necessary in general optimization?}

\noindent\textbf{Q3: Can \emph{PhasePoly} capture long-range optimization opportunities at scale?}

\noindent\textbf{Q4: Do \emph{PhasePoly}'s logical reductions translate to near-term hardware execution improvements?}

\noindent\textbf{Q5: How does \emph{PhasePoly} benefit fault-tolerant compilation pipelines, and how should it be integrated?}

\noindent\textbf{Q6: Is the cross-block optimization correct and robust?}

\noindent\textbf{Q7: What are the compilation cost and parameter sensitivity of \emph{PhasePoly}?}

\subsection{Experiment Setup}
    
    \textbf{Benchmarks.} In our experiment design, we selected a set of benchmark circuits that have been widely used in previous circuit optimization research~\cite{amy2014polynomial, amy2018controlled, hietala2021POPL, kissinger2019pyzx, nam2018npj, xu2022PLDI, xu2023PLDI, li2024OOPSLA, ruiz2025quantum}, complemented by new benchmarks representing near-term and fault-tolerant applications. The suite includes quantum arithmetic circuits, MCX, Hamming coding functions, Hamiltonian simulation, QAOA, Grover's algorithm, and Shor's algorithm.
    
    \textbf{Baselines.} We compare \emph{PhasePoly} with three phase polynomial baselines:  (i) \emph{Rotation Merging}~\cite{nam2018npj} merges phase gates that share identical phase polynomials, but does not optimize the CNOT network.
        (ii) \emph{Single-block Greedy Optimization}~\cite{amy2018controlled,vandaele2022QuantumScienceTechnology} greedily synthesizes the phase-parity network within each phase-polynomial block. The output-parity network is handled separately using Gaussian elimination. 
        (iii) Gray-Synth~\cite{amy2018controlled} reduces two-qubit gates using the \emph{sum-over-paths} representation. We report the CNOT-count results from the original paper.
    
    We also integrate our technique into two general-purpose optimizers. We use Quartz~\cite{xu2022PLDI} and QUESO~\cite{xu2023PLDI}. Quartz performs equivalent \textbf{subcircuit rewriting} and does not model phase contributions. QUESO adopts the \emph{sum-over-paths} form to enhance phase polynomial ECCs and uses search-based rewriting methods to enlarge optimization coverage.
    
    \textbf{Setup.} All experiments run on a 2.8\,GHz AMD EPYC~7313 CPU. \emph{PhasePoly} uses a priority queue and a solution pool with maximum sizes chosen according to the runtime budget. We enable an \emph{Incremental Block Merging} strategy to improve robustness during cross-block optimization. For Quartz and QUESO, we follow their recommended equivalent subcircuit sizes and allocate up to 2-hour per circuit. \emph{PhasePoly} attains the reported results without consuming the full runtime budget.
    
\subsection{Q1: Comparison with Phase Polynomial Baselines}

We compare \emph{PhasePoly} with three phase polynomial baselines, as summarized in Fig.~\ref{fig:table1}.  
    \emph{Rotation Merging} combines only rotation gates with identical phase polynomials and leaves the CNOT network unchanged, which limits the overall optimization gains. We implement this pass using our \emph{SSA-style rotation-merging} infrastructure.
    \emph{Single-block Greedy Optimization} is reproduced in our infrastructure as an independent per-block optimization pass, where the phase-parity and output-parity networks are synthesized separately using greedy synthesis and Gaussian elimination, respectively, without cross-block optimization. This baseline reduces total gates by 26.93\% and two-qubit gates by 8.14\% on average.
        Gray-Synth~\cite{amy2018controlled}, built on T-par~\cite{amy2014polynomial}, targets CNOT reduction; its reported results show an average CNOT reduction of 17.62\%.
        
        In contrast, \emph{PhasePoly} co-optimizes \textbf{the phase-parity network} and \textbf{the output-parity network} and employs \emph{cross-block IR and optimization}. It achieves up to 50\% total-gate reduction and 48.57\% CNOT reduction—\textbf{34.70\%} and \textbf{26.83\%} on average—surpassing all baselines and improving upon Gray-Synth by 9.21\% in CNOT reduction.
    
    \begin{figure*}[t!]
        \vspace{-1\baselineskip}

        \centering
        \includegraphics[width=1.0\textwidth]{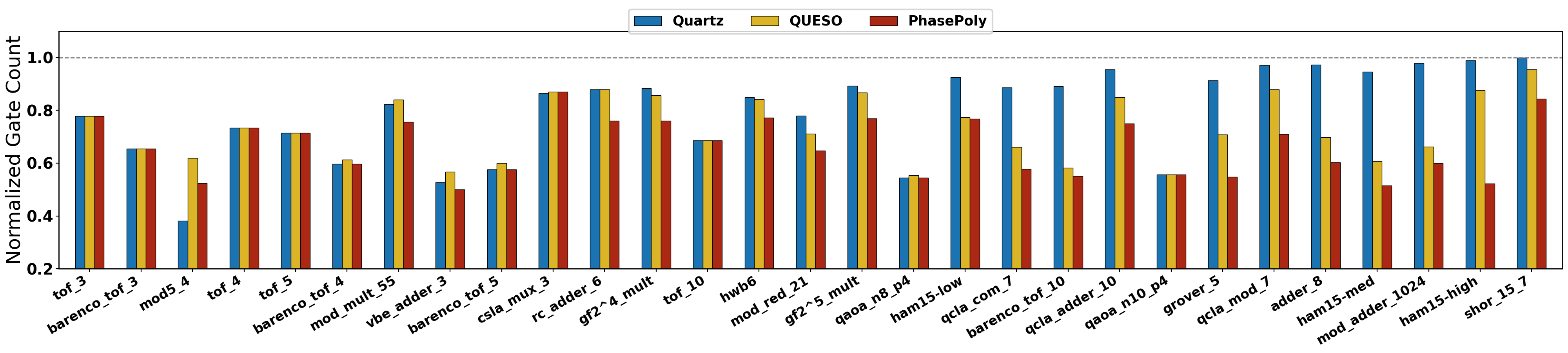}
        \vspace{-1.25\baselineskip}

        \includegraphics[width=1.0\textwidth]{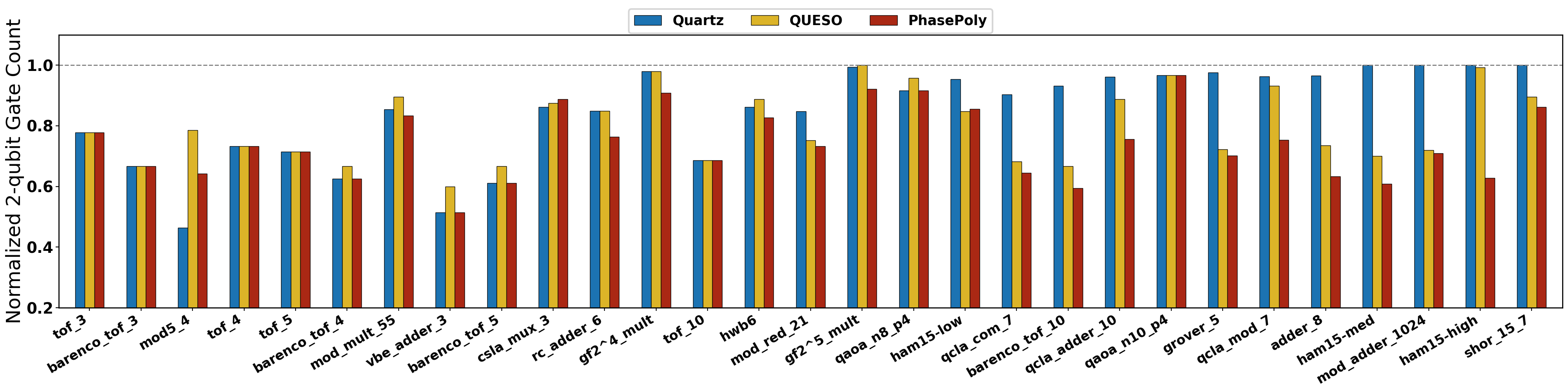}
        \vspace{-1.5\baselineskip}
        \caption{
        Normalized total and two-qubit gate-count reductions across benchmark circuits, comparing \emph{PhasePoly} against general circuit optimizers. All values are normalized to the unoptimized circuits (1.0), with lower bars indicating greater reduction.
        }
        \vspace{-0.5\baselineskip}
        \label{fig:table2}
    \end{figure*}

    \begin{rqsummary}
\textbf{Q1 Summary:} \emph{PhasePoly} outperforms by jointly optimizing phase- and output-parity and exploiting cross-block opportunities missed by phase-only, single-block methods.
    \end{rqsummary}
    \vspace{-0.5\baselineskip}
        
    \subsection{Q2: Necessity in General Circuit Optimization}

    \emph{PhasePoly} is orthogonal to subcircuit rewriting frameworks and can be composed with existing optimization passes. 
    To study their interaction, we integrate \emph{PhasePoly} with two state-of-the-art subcircuit rewriting frameworks—Quartz~\cite{xu2022PLDI} and QUESO~\cite{xu2023PLDI}—and evaluate them under their recommended settings (3-qubit / 6-gate subcircuits, 2-hour per circuit).
    Fig.~\ref{fig:table2} summarizes their standalone performance; note that this is not an apples-to-apples comparison: although they optimize gates beyond phase polynomial structure, \emph{PhasePoly} still delivers the strongest reductions: Quartz and QUESO reduce total gate by 22.17\% and 27.83\% (CNOTs by 16.88\% and 20.70\%), while \emph{PhasePoly} achieves total reduction by \textbf{34.70\%} (CNOTs by \textbf{26.83\%}) on average.
    
    \paragraph{Effect of circuit scale}
    On average, \emph{PhasePoly} achieves greater reductions in both total and CNOT gates, though it is not always the best on small circuits. 
    To analyze this trend, we group benchmarks by original gate count:
    
    \textbf{Small ($<$200 gates):} Taking the best of Quartz and QUESO per circuit, \emph{subcircuit rewriting} ties \emph{PhasePoly} on 6 circuits and surpasses it on 2 of 10.
    
    \textbf{Medium (200--500 gates):} Only 3 circuits tie or exceed \emph{PhasePoly} among 10.
    
    \textbf{Large ($>$500 gates):} Only one QAOA circuit ties \emph{PhasePoly}; all other circuits exhibit significant performance gaps.
    
    These results are consistent with design intent: \emph{PhasePoly} leverages the \emph{phase polynomial} and \emph{cross-block} IR to capture long-range structure across the circuit, while subcircuit rewriting—bounded by local equivalence patterns—loses effectiveness as it scales up despite covering more gate types.
    
\begin{figure}[t!]

        \centering
        \includegraphics[width=0.49\columnwidth]{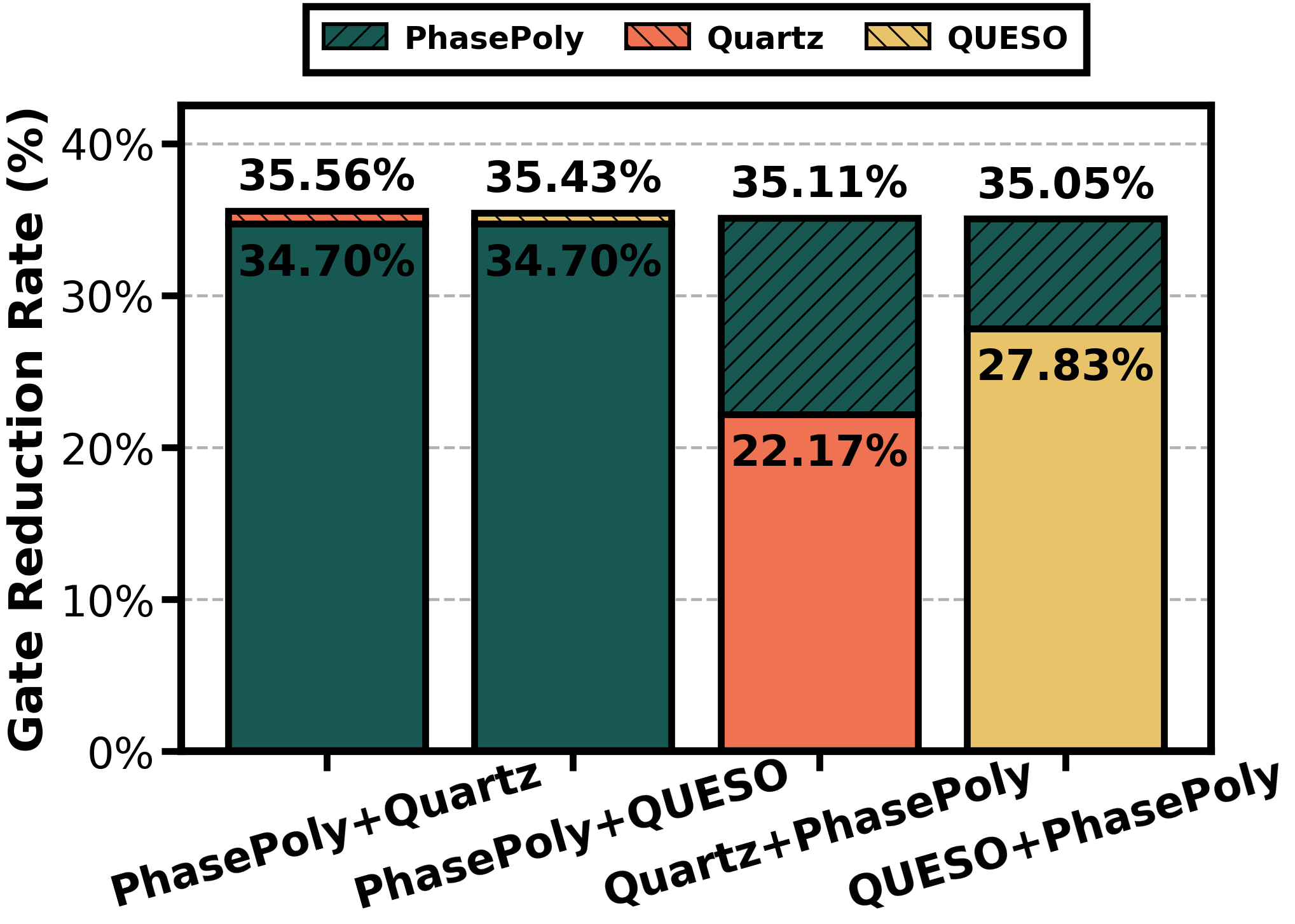}
        \hfill
        \includegraphics[width=0.49\columnwidth]{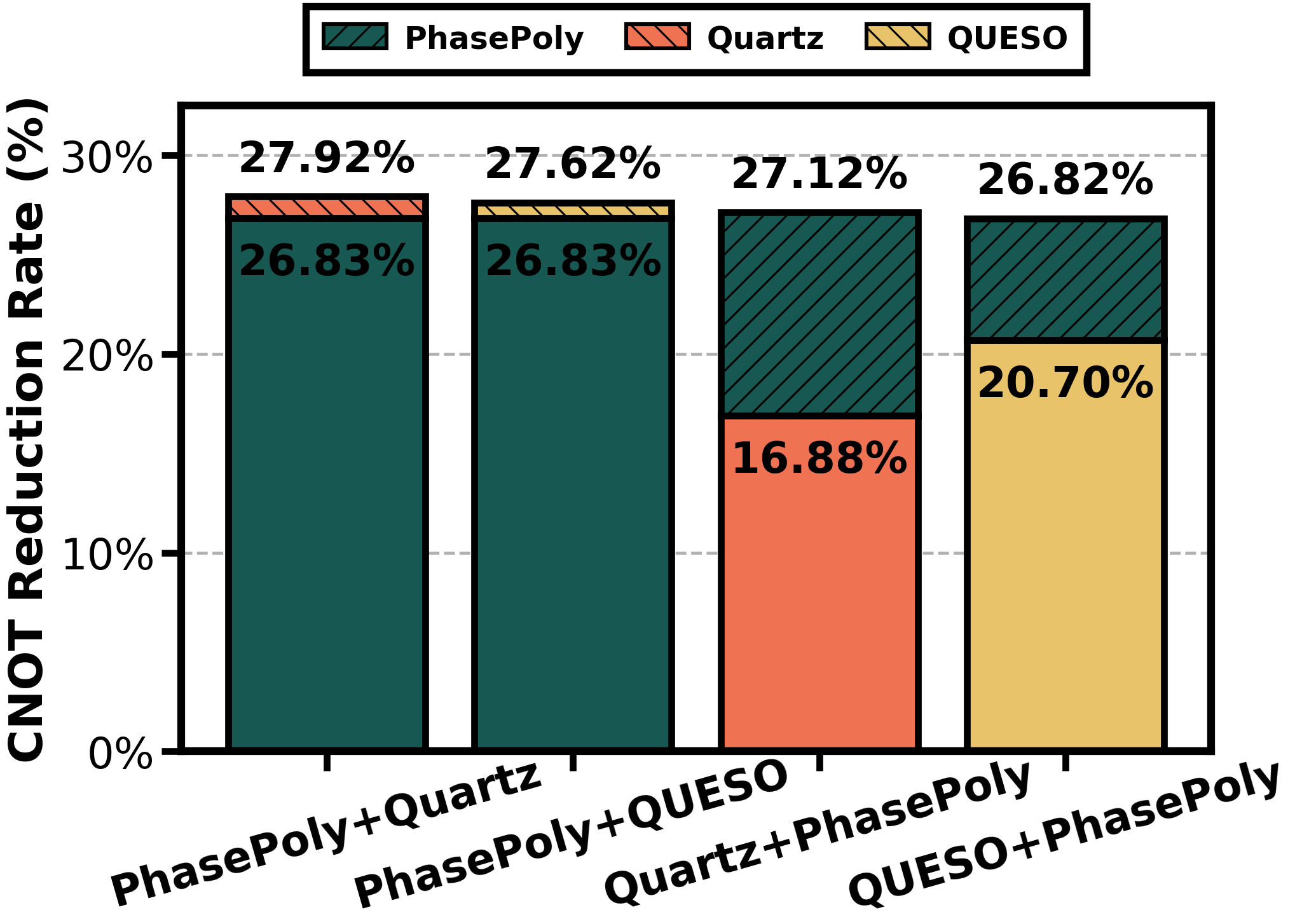}
    
        \caption{
            Integration of \emph{PhasePoly} with Quartz and QUESO. 
    Bars show average total-gate (left) and CNOT (right) reduction rates relative to original circuits. 
    “A+B” applies A then B; solid and hatched bars indicate first and second passes. 
        }
        \vspace{-1.0\baselineskip}
        \label{fig:collaboration}
    \end{figure}

    \paragraph{Integrating optimization passes}
    Fig.~\ref{fig:collaboration} analyzes how \emph{PhasePoly} interacts with subcircuit rewriting in a combined pipeline. 
    We denote sequential application as ``A+B'' (run A first, then B). 
    Applying subcircuit rewriting after \emph{PhasePoly} yields modest additional gains ($\approx$0.75--1.25\% in total and CNOT reductions). 
    Applying \emph{PhasePoly} after rewriting provides substantially larger improvements ($\approx$6--13\%), revealing that \emph{PhasePoly} identifies long-range opportunities left unexploited by local subcircuit rewriting.  
        
    Although QUESO slightly outperforms Quartz as a standalone pass—thanks to its phase modeling—both benefit when combined with \emph{PhasePoly}. 
     Quartz + \emph{PhasePoly} performs better than QUESO + \emph{PhasePoly}. Quartz does not consider any phase information, and therefore mainly removes local redundancy. This implies the benefit of having a dedicated phase polynomial pass, rather than having the phase polynomial optimization spread into different optimization passes. 
    
    \vspace{-0.5\baselineskip}
    \begin{rqsummary}
        \textbf{Q2 Summary:} \emph{PhasePoly} complements local subcircuit rewriting: the former enables long-range CNOT/phase reductions, while the latter is effective on small circuits. Together, they unlock optimizations neither achieves alone.
    \end{rqsummary}
    \vspace{-0.5\baselineskip}

    \subsection{Q3: Can {PhasePoly} Effectively Scale to Large Circuits?}

Q2 shows that the gap between \emph{PhasePoly} and subcircuit rewriting widens as circuits grow. Search-based \emph{subcircuit rewriting} enumerates many candidate equivalent subcircuits before applying a rule. As the target pattern size grows, the search space explodes exponentially, so practical deployments restrict patterns to limited windows. This locality makes it difficult to realize long-range optimization opportunities that span large portions of a circuit.

    We further stress-test scalability on three large-circuit families:
    \emph{(i) MCX} (multi-controlled-NOT) circuits~\cite{barenco1995elementary}, where qubit and gate counts grow roughly linearly (19--499 qubits; 480--14{,}880 gates);
    \emph{(ii) Adder} circuits (23--383 qubits; 637--12{,}637 gates); and
    \emph{(iii) HWB} (Hamming coding functions)~\cite{saeedi2010reversible} with a fixed count of 16 qubits but rapidly growing gates (345--104{,}068).
    We compare \emph{PhasePoly} with Quartz and QUESO under a 2-hour time budget; \emph{PhasePoly} never exceeded 5{,}500\,seconds even on the largest instance (\texttt{hwb8\_113}).

\begin{figure}[htbp]
            \vspace{-1\baselineskip}

    \centering
    \begin{subfigure}{\linewidth}
        \centering
        \includegraphics[width=0.49\linewidth]{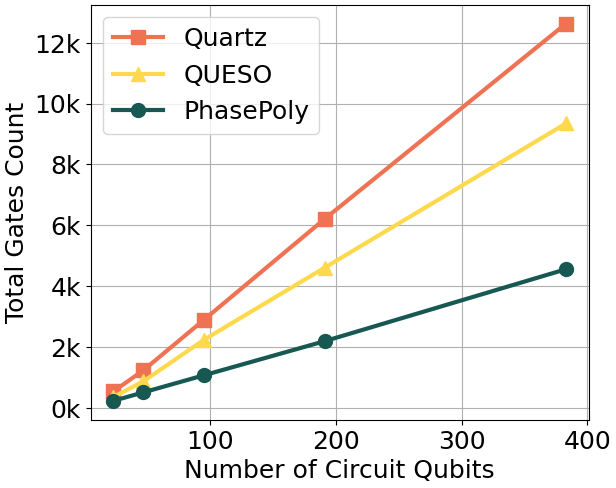}
        \hfill
        \includegraphics[width=0.48\linewidth]{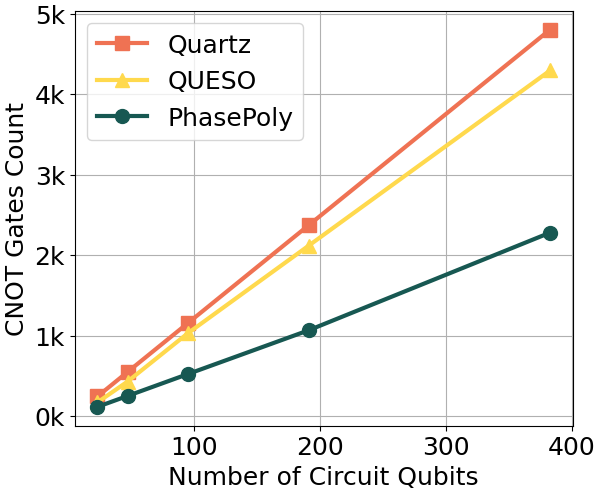}
        \vspace{-0.5em}
    \caption{Adder circuits. X-axis: qubits; Y-axis: total gates and CNOTs.}    \label{fig:large_circuit_adder}
    \end{subfigure}
    
    \begin{subfigure}{\linewidth}
        \centering
        \includegraphics[width=0.49\linewidth]{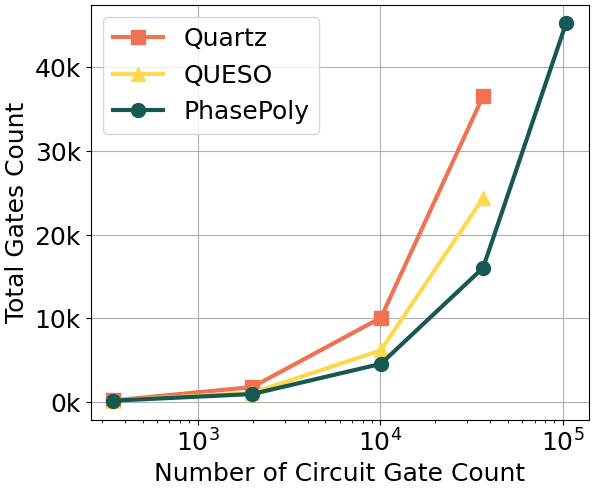}
        \hfill
        \includegraphics[width=0.49\linewidth]{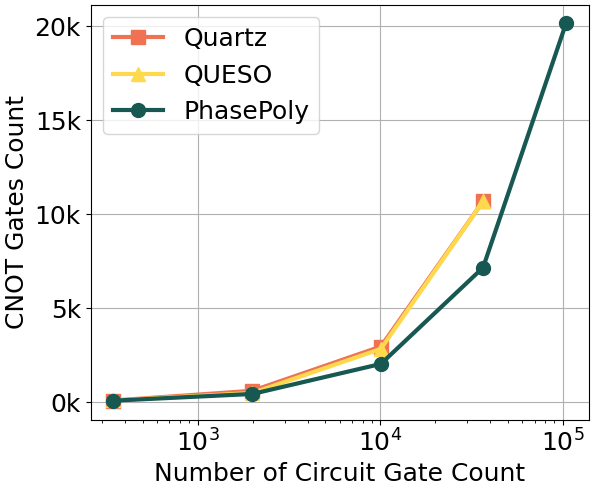}
           
        \caption{HWB circuits. X-axis: original gate count (log scale); Y-axis: total gates and CNOTs. Missing points denote optimization failures.}
        \label{fig:large_circuit_hwb}
    \end{subfigure}
        \vspace{-1.5em}
        \caption{
        Comparison of large circuits,  
        \emph{PhasePoly} scales effectively, while subcircuit rewriting saturates on large circuits.
    }
    \vspace{-0.5\baselineskip}
    \label{fig:large_circuit_comparison}
    \end{figure}

    Fig.~\ref{fig:large_circuit_comparison} reports total-gate and CNOT reductions for adder and HWB circuits, complementing the MCX results in Fig.~\ref{fig:motivation_mcx}. 
    As circuit size increases, Quartz and QUESO saturate or fail on the largest instances, whereas \emph{PhasePoly} continues to achieve substantial reductions. 
    These results show that structured parity-matrix reasoning and cross-block optimization expose long-range opportunities that fixed-window subcircuit rewriting struggles to capture.
    
    \vspace{-0.5\baselineskip}
    \begin{rqsummary}
    \textbf{Q3 Summary:} On large circuits, \emph{PhasePoly} sustains and widens its advantage by leveraging its global parity reasoning and cross-block optimization, capturing long-range reductions that \emph{subcircuit rewriting} cannot.
    \end{rqsummary}
    \vspace{-0.5\baselineskip}

    \subsection{Q4: Hardware-Level Execution Improvements}
    \label{sec:hardware_level_execution}
    On near-term hardware, execution quality is often limited by (i) circuit depth (accumulated error) and (ii) constrained connectivity, where non-local two-qubit interactions introduce routing overhead (SWAPs). Therefore, gate-count reductions must be validated against depth and physical circuit metrics.

\begin{figure}[htbp]
        \vspace{-0.5\baselineskip}
        \centering
        \begin{subfigure}{\linewidth}
            \centering
            \includegraphics[width=0.98\linewidth]{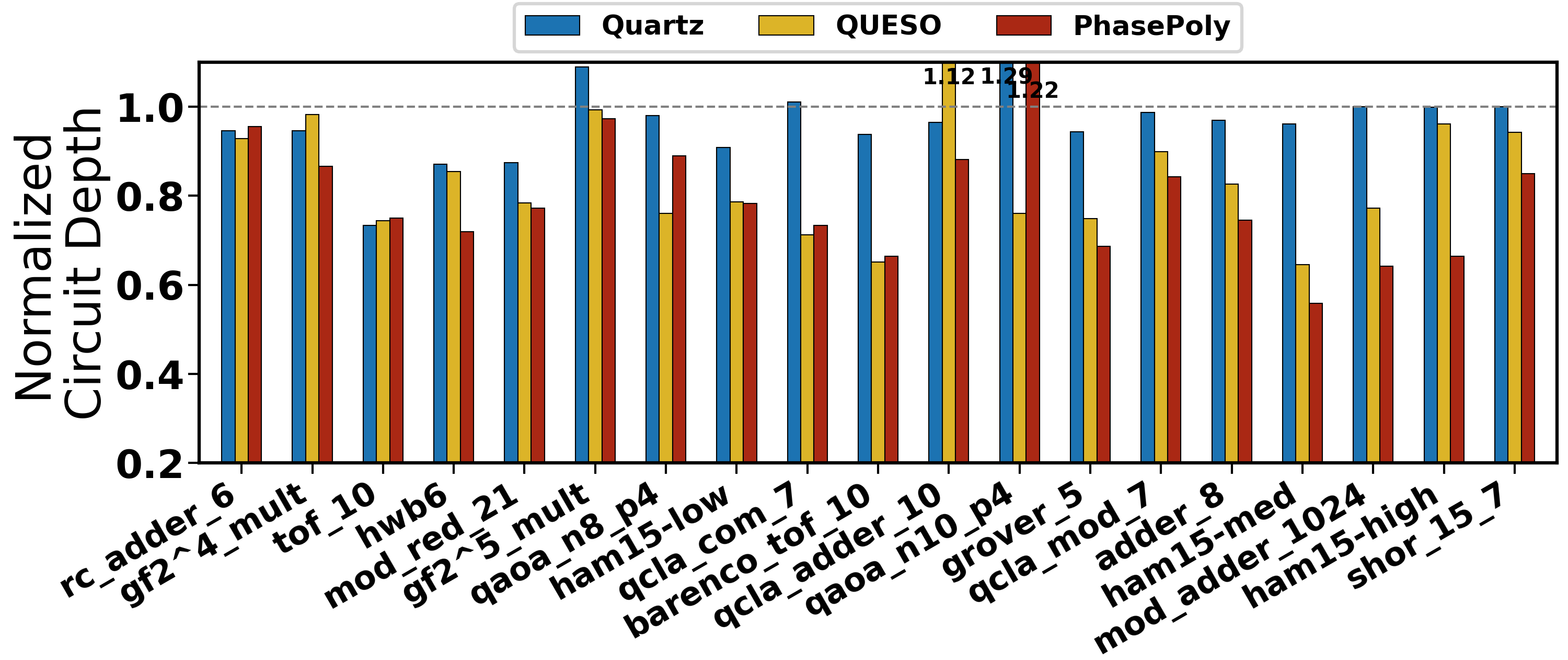}
            \vspace{-0.5em}
            \caption{Normalized circuit depth reductions across benchmark circuits.}
            \label{fig:depth_study_benchmark}
        \end{subfigure}
        
        \begin{subfigure}{\linewidth}
            \centering
            \includegraphics[width=0.31\linewidth]{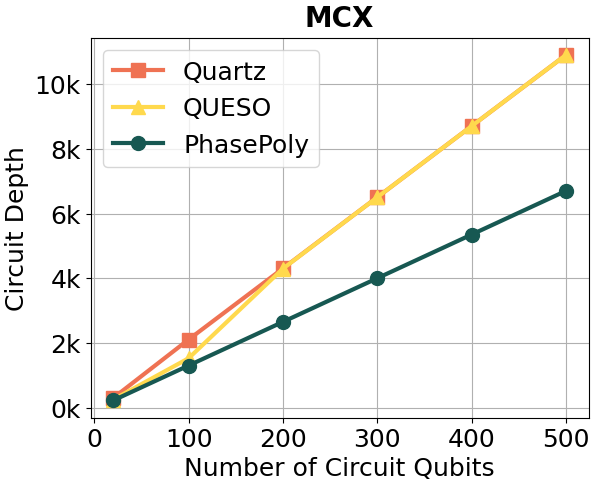}\hspace{0.5em}%
            \includegraphics[width=0.32\linewidth]{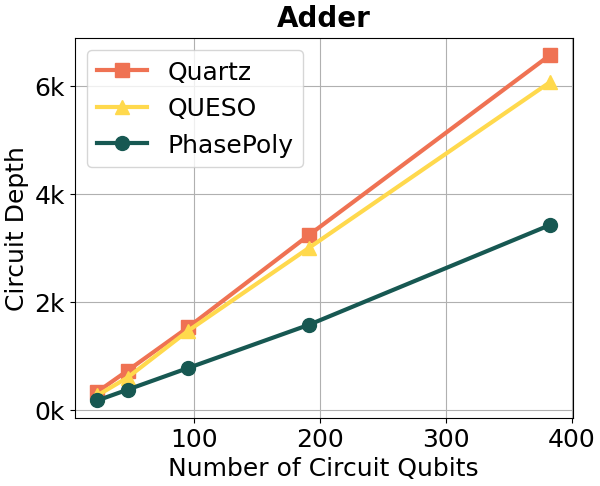}\hspace{0.5em}%
            \includegraphics[width=0.31\linewidth]{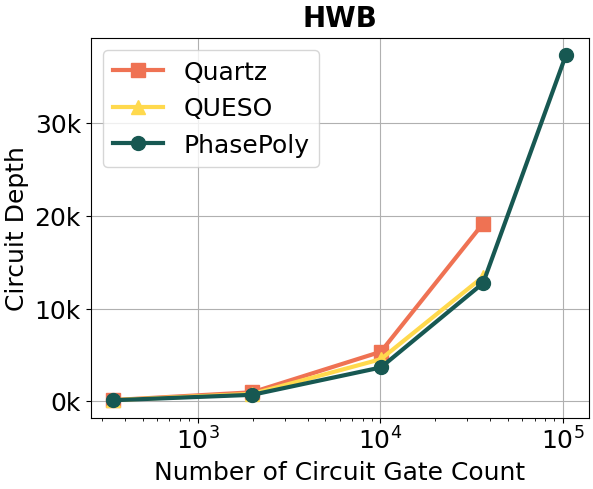}
            \vspace{-0.5em}
            \caption{Normalized circuit depth reductions for MCX, Adder, and HWB circuit families. The HWB x-axis is shown in log scale.}
            \label{fig:depth_study_large}
        \end{subfigure}
        
        \caption{Normalized logical circuit depth reductions across benchmark circuits and three large circuit families.}  
        
        \vspace{-0.5\baselineskip}
        \label{fig:depth_study_combined}
    \end{figure}

    \textbf{Logical depth.} Fig.~\ref{fig:depth_study_combined}(a) reports normalized logical circuit depth (excluding small-size circuits for readability). On average, Quartz reduces depth by 13.26\%, QUESO by 19.64\%, and \emph{PhasePoly} achieves the largest reduction of 22.47\%, corresponding to a $1.14$--$1.69\times$ improvement.
    The gap widens on large circuit families in Fig.~\ref{fig:depth_study_combined}(b), where Quartz reduces depth by 4.03\%, QUESO by 14.45\%, and \emph{PhasePoly} by 40.91\%, corresponding to a $2.83$--$10.15\times$ improvement. 
    
        \textbf{Hardware mapping under limited connectivity.} To test whether logical improvements persist after routing, we map optimized circuits to 2D planar coupling graphs (square grids) and perform routing using Qiskit SABRE~\cite{gadi_aleksandrowicz_2019_zenodo, li+:asplos19, zou2024lightsabre}. We report (i) \emph{weighted two-qubit} gate count and (ii) \emph{physical circuit depth}, where each SWAP is weighted as three CNOTs.

\begin{figure}[t!]
        \centering
    
    \begin{subfigure}{\linewidth}
    \vspace{-1\baselineskip}

        \centering
    
        \includegraphics[width=0.98\textwidth]{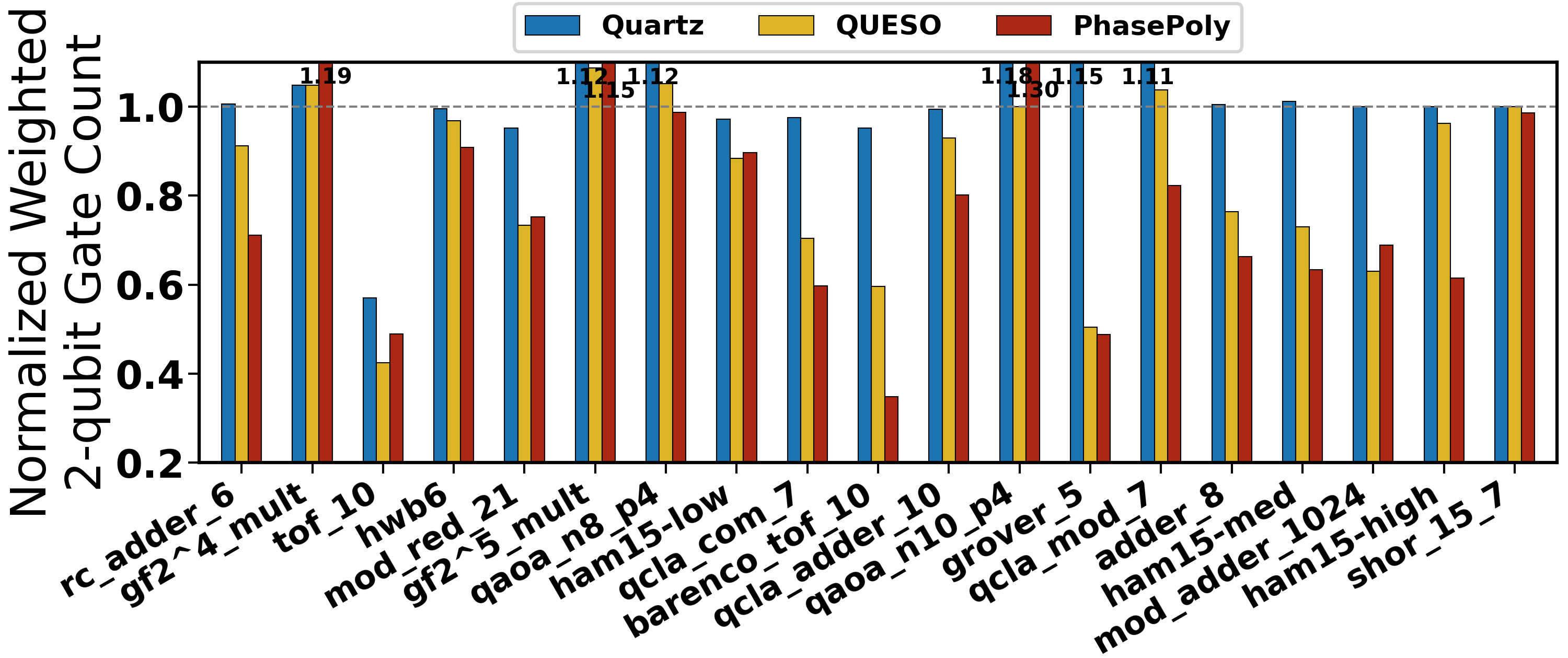}
    
    
        \includegraphics[width=0.98\textwidth]{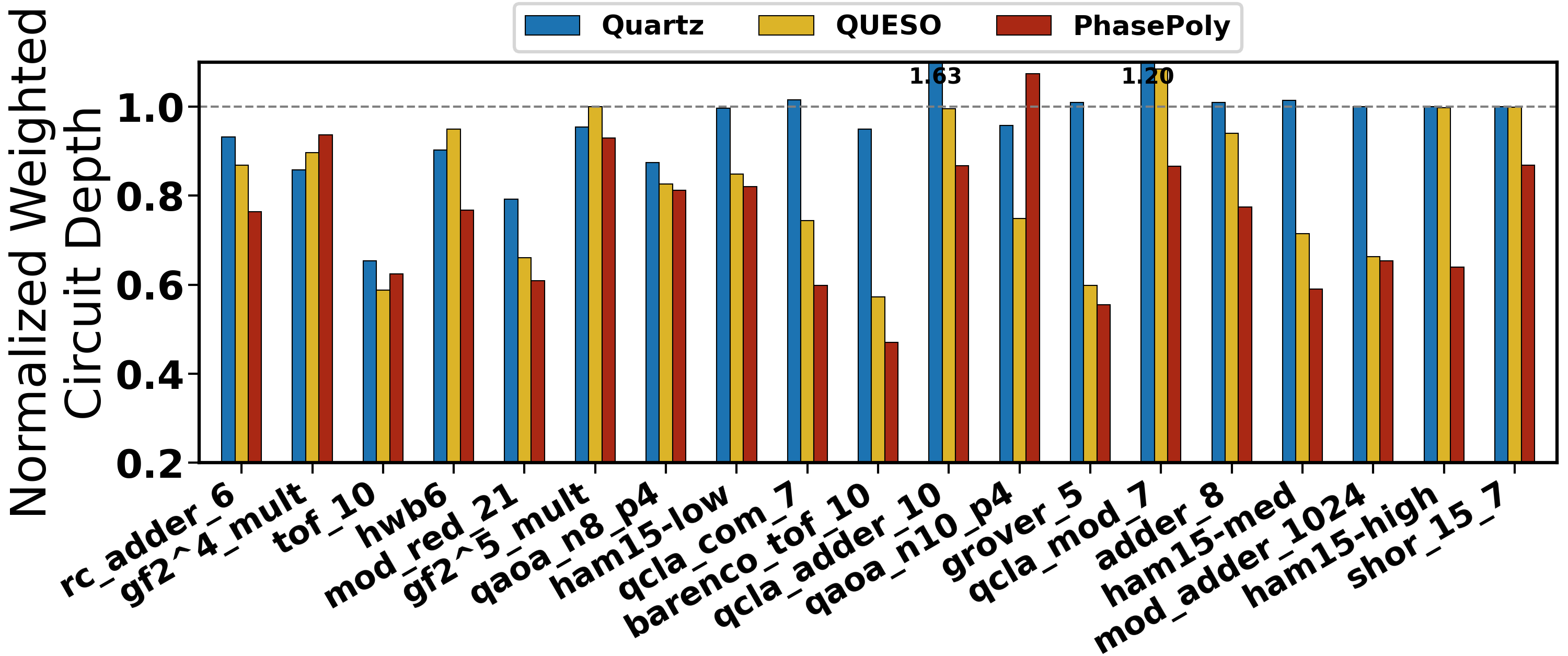}
        \vspace{-0.5em}
        \caption{
        Normalized weighted two-qubit gate count and physical circuit depth across benchmark circuits.
        }
        \label{fig:mapping_study_benchmark}
    \end{subfigure}
    
    \begin{subfigure}{\linewidth}
        \centering
        \includegraphics[width=0.31\linewidth]{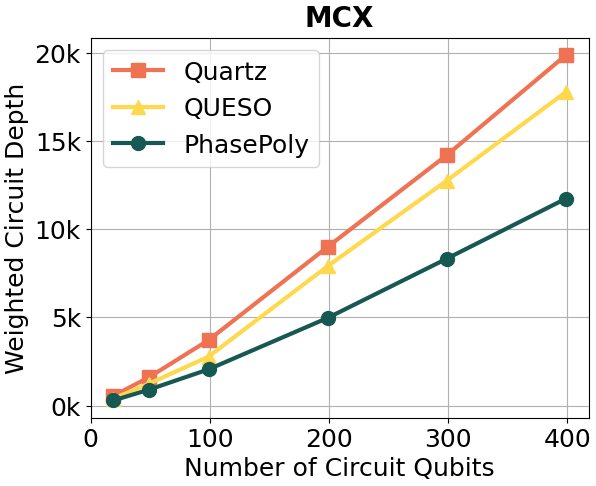}\hspace{0.5em}%
        \includegraphics[width=0.32\linewidth]{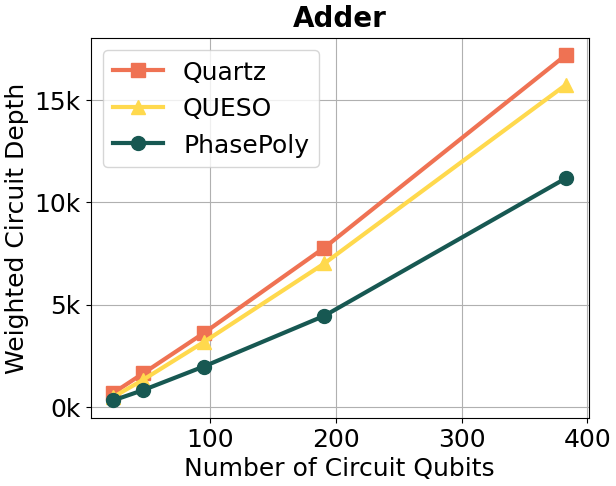}\hspace{0.5em}%
        \includegraphics[width=0.31\linewidth]{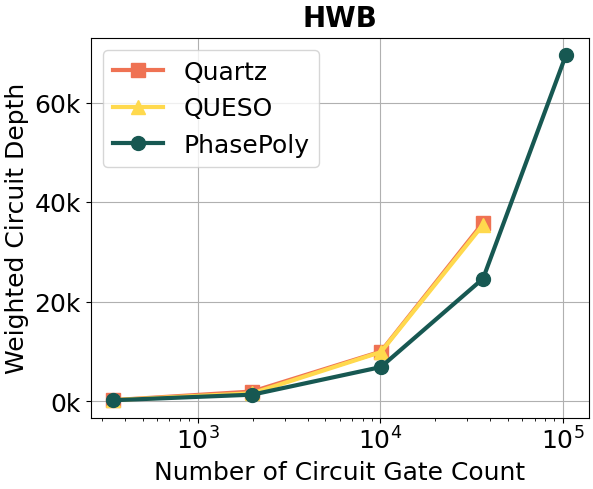}
        \vspace{-0.5em}
        \caption{Weighted physical circuit depth for MCX, Adder, and HWB circuit families. The HWB x-axis is shown in log scale.}
        \label{fig:physical_depth_study_large}
    \end{subfigure}
    \vspace{-1\baselineskip}
    \caption{Reductions in weighted physical circuit 2-qubit gate-count and depth after hardware mapping.}
    \label{fig:physical_depth_study}
    \vspace{-1\baselineskip}
    \end{figure}

        Fig.~\ref{fig:physical_depth_study}(a) shows that, across benchmark circuits, on average, Quartz reduces physical circuit depth by 15.23\%, QUESO by  21.60\%, while \emph{PhasePoly} achieves the largest reductions of 28.35\%, corresponding to a $1.31$--$1.86\times$ improvement in physical circuit depth.
        On large circuit families (Fig.~\ref{fig:physical_depth_study}(b)), the advantage becomes stronger: Quartz reduces physical depth by 2.7\%, QUESO by 15.25\%, and \emph{PhasePoly} by 40.84\%, corresponding to a $2.68$--$15.13\times$ improvement.
    
        \textbf{Why hardware mapping can amplify circuit optimization gains (and when it does not).} In many cases, \emph{PhasePoly}'s logical reductions \emph{amplify} after mapping (e.g., depth reductions of 22.47\% become 28.35\% in physical circuits), because fewer two-qubit interactions lead to fewer SWAPs. Cross-block optimization further helps by reusing parity structures across phase-polynomial blocks, reducing repeated reconstruction of multi-qubit interactions. \emph{PhasePoly} matches or outperforms Quartz and QUESO on most benchmarks. However, for \texttt{qaoa\_n10\_p4}, CNOT count decreases by $\sim$3\% but depth increases by $\sim$20\% due to gate-count optimization, potentially reducing parallelism. For QAOA-type circuits, domain-specific hardware mappers \cite{alam+:dac20, jin+:asplos23} exist and exploit commuting flexibilities unique to QAOA circuits. 
        We expect that the domain-specific hardware mappers may further reduce gate counts on top of already optimized logical circuits. 
        
        For circuits that are already topology-friendly (e.g., \texttt{gf2\textasciicircum4\_mult}, \texttt{gf2\textasciicircum5\_mult}), routing overhead can dominate and offset logical gate-count gains (all optimized circuits have worse physical performance than original circuits). Overall, for the majority of cases, as circuits scale, \emph{PhasePoly}'s reductions translate reliably into better physical circuit depth and two-qubit gate cost.

    \vspace{-0.25\baselineskip}
    \begin{rqsummary}
        \textbf{Q4 Summary:} \emph{PhasePoly} reduces not only gate counts but also logical depth. Under constrained connectivity, these gains persist---and often amplify---after routing, yielding substantial reductions in depth and two-qubit gate cost, especially for large circuits.
    \end{rqsummary}
    \vspace{-0.5\baselineskip}
    
    \subsection{Q5: Fault-Tolerant Compilation Benefits and Integration}
    \label{sec:ft_compilation}

    Fault-tolerant quantum computing (FTQC) is limited by qubit overhead, runtime, and architectural constraints. 
    Many logical/NISQ optimizations extend to FT compilation~\cite{forster2025quantum}. 
    Traditional FT analyses focus on $T$-gate count/depth due to the high cost of magic-state distillation~\cite{bravyi2005universal}. 
    Recent advances in magic-state cultivation~\cite{gidney2024magic,chen2025efficient, rosenfeld2025magic} substantially reduce this overhead, making Clifford costs increasingly important. 
    Modern cost models suggest that CNOTs can be \textbf{comparable} in spacetime cost to $T$ states of similar reliability~\cite{gidney2024magic}, with \textbf{non-constant} ancilla volume and operation depth~\cite{huggins2025fluid}.
    Since \emph{PhasePoly} reduces both CNOT and $R_z$ structure at the logical level, it can improve downstream FT resource costs.

\begin{figure}[htbp]
        \vspace{-0.5\baselineskip}
        \centering\includegraphics[width=0.48\textwidth]{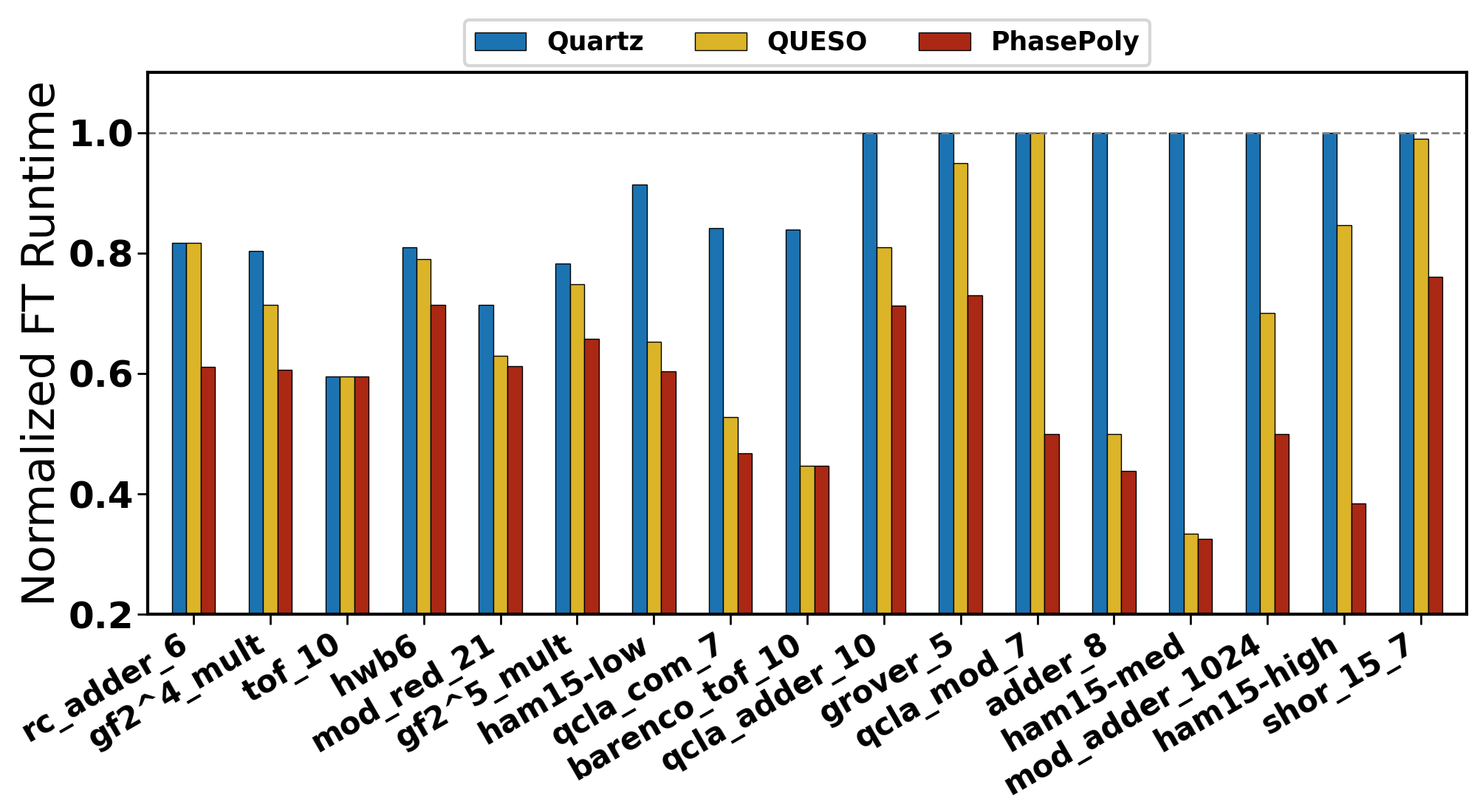}
        \vspace{-1em}
        \caption{Normalized fault-tolerant circuit wall-clock runtime.}
        \label{fig:ft-runtime}
    \end{figure}

\textbf{FT resource estimation.}
    Using the Azure Resource Estimator~\cite{beverland2022assessing, van2023using} under a surface-code, nearest-neighbour architecture~\cite{wang2010quantum,fowler2012surface}, we perform end-to-end FT resource estimation. Fig.~\ref{fig:ft-runtime} reports normalized wall-clock runtime relative to unoptimized circuits, excluding small-size and parameterized circuits.  
    Quartz, QUESO, and \emph{PhasePoly} achieve average reductions of 11.99\%, 31.80\%, and 44.62\%, respectively, with \emph{PhasePoly} providing the largest improvement.

    \textbf{Integration with Clifford+$T$ synthesis.}
    Each arbitrary $R_z$ rotation must be synthesized into an FT instruction set such as Clifford+$T$ ($H$, $S$, and $T$ sequence). We therefore study how \emph{PhasePoly} interacts with FT gate synthesis by combining it with GridSynth~\cite{ross2016optimal} on 14 variational circuits: QAOA Max-Cut on 3-regular graphs (4--24 qubits, 2{,}150--12{,}900 gates)
    and VQE circuits, including UCCSD ansatz with Jordan-Wigner (JW)~\cite{jordan1928paulische},
    Bravyi-Kitaev (BK)~\cite{bravyi2002fermionic}, and parity (P)~\cite{seeley2012bravyi}
    encodings, as well as the Hamming-weight-preserving ansatz (HW)~\cite{monbroussou2025trainability}
    (4--12 qubits, 2{,}641--231{,}780 gates). We compare two compilation orders:
    \textbf{(A) GridSynth $\rightarrow$ \emph{PhasePoly}} and
    \textbf{(B) \emph{PhasePoly} $\rightarrow$ GridSynth}. The purpose of investigating this is to see how to better apply \emph{PhasePoly} into the overall compilation pipeline. 
    Both pipelines apply the same commuting-rule simplification as the final pass.

    \begin{figure}[t!]
        \centering
        \vspace{-1\baselineskip}
        \begin{subfigure}{\linewidth}
            \centering
            \includegraphics[width=0.98\textwidth]{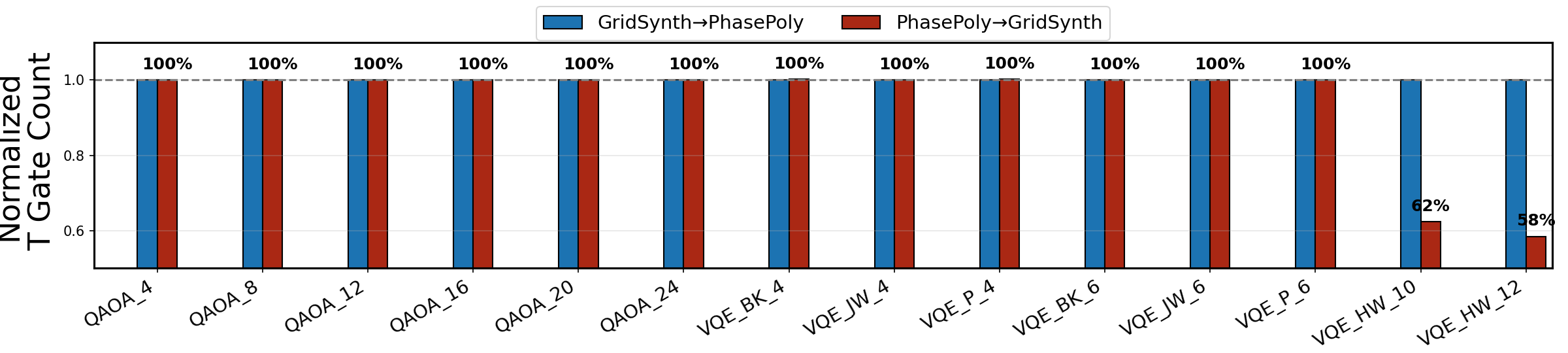}
            \vspace{-0.5em}
            \caption{Normalized $T$-gate count.}
            \label{fig:gridsynth_t}
        \end{subfigure}
        
        \begin{subfigure}{\linewidth}
            \centering
            \includegraphics[width=0.98\textwidth]{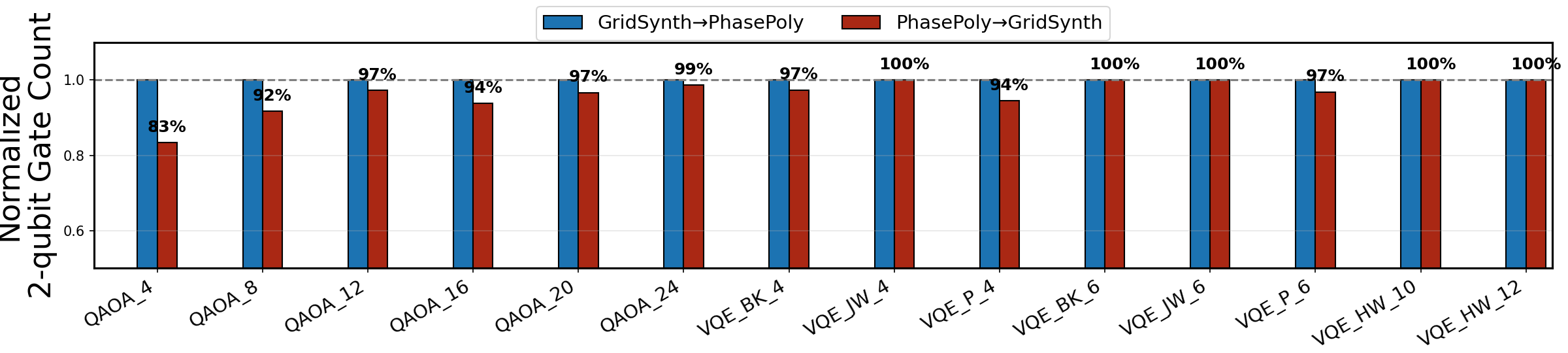}
            \vspace{-0.5em}
            \caption{Normalized two-qubit gate count.}
            \label{fig:gridsynth_cx}
        \end{subfigure}
        
        \begin{subfigure}{\linewidth}
            \centering
            \includegraphics[width=0.98\textwidth]{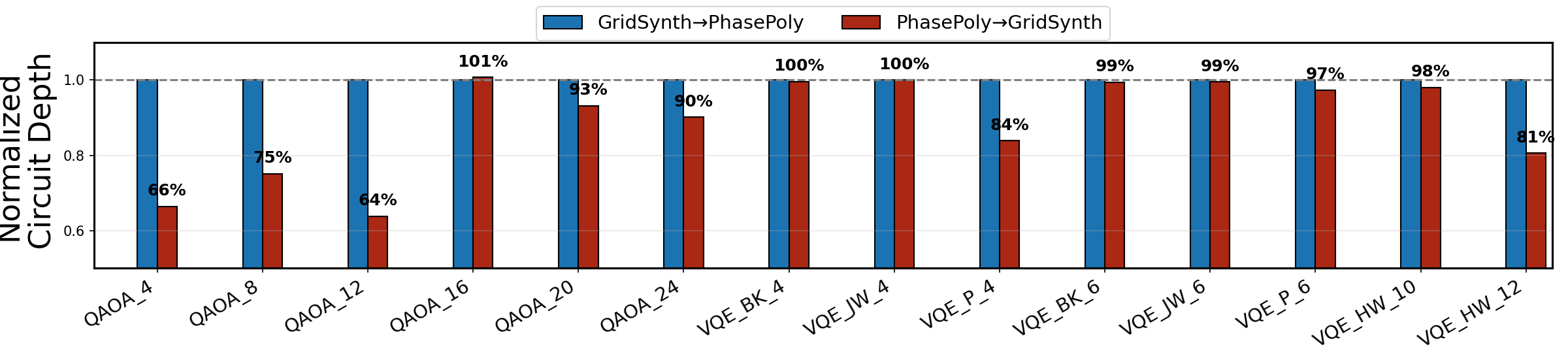}
            \vspace{-0.5em}
            \caption{Normalized circuit depth.}
            \label{fig:gridsynth_depth}
        \end{subfigure}
        
        \vspace{-0.5em}
    \caption{Comparison of two compilation orders. Metrics are normalized to GridSynth $\rightarrow$ \emph{PhasePoly}.}
        \vspace{-1.0\baselineskip}
        \label{fig:gridsynth_study}
        
        \end{figure}    

        Fig.~\ref{fig:gridsynth_study} reports normalized $T$ count,
        two-qubit gate count, and circuit depth.
        Running \emph{PhasePoly} before GridSynth produces the lowest depth on most circuits because large \{CNOT,$R_z$\} regions are simplified before GridSynth
        introduces additional $H$ gates that split phase-polynomial blocks
        and limit rotation-merging opportunities.
        
        Across benchmarks, we observe:
        (i) \textbf{$T$-count changes are modest}, noticeable mainly for HWPA circuits;
        (ii) \textbf{two-qubit gate reductions are common} using \emph{PhasePoly}'s optimization and often translate to lower depth;
        and (iii) \textbf{the benefit depends on the regularity of circuit structure}: circuits with structured
        phase interactions (e.g., parity and HWPA ansatz) obtain larger improvements ($\sim$10\% depth reduction on average), while JW/BK encodings show $<1\%$ change due to their already compact CNOT-$R_z$ structure.

    \vspace{-0.5\baselineskip}
    \begin{rqsummary}
    \textbf{Q5 Summary:}
    \emph{PhasePoly}'s strong CNOT/$R_z$ reductions, long-range optimization, and natural fit for Clifford+$T$ circuits make it effective for fault-tolerant compilation.
    \emph{PhasePoly} is most effective when applied before Clifford+$T$ synthesis, which introduces additional $H$ barriers.
    \end{rqsummary}
    \vspace{-0.5\baselineskip}
    
    \subsection{Q6: Correctness and Robustness of {PhasePoly}} 
    \label{sec:rq4}
    
    \paragraph{Equivalence checking for correctness}
    Phase polynomials faithfully model \{CNOT, $R_z$\} circuits and are widely used for verification~\cite{amy2017verified,amy2018towards,xu2023PLDI}. Because \emph{PhasePoly} introduces \emph{cross-block IR and optimization} that merges multiple phase-polynomial blocks, we enforce additional constraints and pruning rules to guarantee that all intermediate states remain legal and synthesizable (Section~\ref{sec:cross-block-ir-linear}).
    
        We also perform end-to-end equivalence checking: for circuits with fewer than 8 qubits, we compare unitaries using Qiskit~\cite{gadi_aleksandrowicz_2019_zenodo}, and for all circuits we use the formal verification tool MQT~QCEC~\cite{burgholzer2020advanced}. All checked benchmarks pass verification; \texttt{mod\_adder\_1024} is excluded because it exceeds our hardware limits.
    
    \paragraph{Incremental block merging for robustness}
    Cross-block optimization can yield additional reductions—about one third of benchmarks (9 circuits) benefit from it. We use \emph{incremental block merging} that expands the merge size gradually, optimizing step by step rather than merging all blocks at once.
    
    Table~\ref{tab:table3} evaluates three representative circuits under the same parameters except for the merge size (“Group~$k$”, merging 1-7 adjacent blocks) and our \emph{Incremental} strategy, which gradually increases $k$ and keeps only beneficial steps. 
        For \texttt{barenco\_tof\_10}, improvements emerge at Group~3 and stabilize, matching the \emph{Incremental} result. 
        For \texttt{adder\_8}, larger groups continue to help (CX: 274 $\rightarrow$ 257), while \emph{Incremental} remains close (259). 
        For \texttt{ham15\_med}, performance peaks at Group~5 (353 $\rightarrow$ 350) but degrades at Group~7; \emph{Incremental} avoids this and achieves the best (325). 
        Overall, \emph{Incremental Block Merging} offers a robust approach that captures large gains while avoiding over-merging regressions.
    
    \vspace{-0.5\baselineskip}
    \begin{rqsummary}
    \textbf{Q6 Summary:} Our \emph{cross-block} IR and optimization preserve correctness (all verified except one timeout) and \emph{Incremental block merging} strategy yields robust gains while avoiding over-merging side effects.
    \end{rqsummary}
    \vspace{-0.5\baselineskip}

        \begin{table}[t!]
            \centering
                    \vspace{-1\baselineskip}
            \renewcommand{\arraystretch}{0.8}
            \resizebox{0.45\textwidth}{!}{
        
        \begin{tabular}{|c|cc|cc|cc|}
        
        \hline
        \multirow{2}{*}{\textbf{Circuit}} & \multicolumn{2}{c|}{\texttt{barenco\_tof\_10}}          & \multicolumn{2}{c|}{\texttt{adder\_8}}       & \multicolumn{2}{c|}{\texttt{ham15\_med}}          \\ \cline{2-7} 
                                 & \multicolumn{1}{c|}{\# Gates} & \# CXs & \multicolumn{1}{c|}{\# Gates} & \# CXs & \multicolumn{1}{c|}{\# Gates} & \# CXs \\ \hline
        Org.                     & \multicolumn{1}{c|}{450}      & 192    & \multicolumn{1}{c|}{900}      & 409    & \multicolumn{1}{c|}{1272}      & 534    \\ \hline
        Group 1                  & \multicolumn{1}{c|}{262}      & 128    & \multicolumn{1}{c|}{557}      & 274    & \multicolumn{1}{c|}{696}      & 353    \\ \hline
        Group 3                  & \multicolumn{1}{c|}{\textbf{248}}      & \textbf{114}    & \multicolumn{1}{c|}{542}      & 259    & \multicolumn{1}{c|}{695}      & 352    \\ \hline
        Group 5                  & \multicolumn{1}{c|}{\textbf{248}}      & \textbf{114}    & \multicolumn{1}{c|}{\textbf{540}}      & \textbf{257}    & \multicolumn{1}{c|}{693}      & 350    \\ \hline
        Group 7                  & \multicolumn{1}{c|}{\textbf{248}}      & \textbf{114}   & \multicolumn{1}{c|}{\textbf{540}}      & \textbf{257}     & \multicolumn{1}{c|}{694}      & 351    \\ \hline
        \emph{Incremental}                & \multicolumn{1}{c|}{\textbf{248}}      & \textbf{114}      & \multicolumn{1}{c|}{542}      & 259     & \multicolumn{1}{c|}{\textbf{656}}      & \textbf{325}    \\ \hline
        \end{tabular}}
        
        \caption{        
            Effect of cross-block merge size on optimization quality for three typical circuits.
        ``Group~$k$'' merges $k$ adjacent blocks at a time, $k\in\{1,3,5,7\}$; ``\emph{Incremental}'' increases $k$ stepwise up to 7, retaining gains and pruning regressions.
        \textbf{Bold} numbers denote the best value in each column.
        }
                \vspace{-1\baselineskip}
        \label{tab:table3}
    \end{table}
    
    \subsection{Q7: Compilation Cost and Parameter Sensitivity} 
    \label{sec:rq5}

\emph{PhasePoly} expands the optimization space beyond greedy phase-parity synthesis by jointly optimizing phase and output parities and enabling cross-block merging. 
We therefore evaluate two practical questions: how quickly it converges, and how sensitive it is to the search parameters.

\begin{figure}[htbp]
    \vspace{-0.5\baselineskip}
        \centering
        \includegraphics[width=0.4\textwidth]{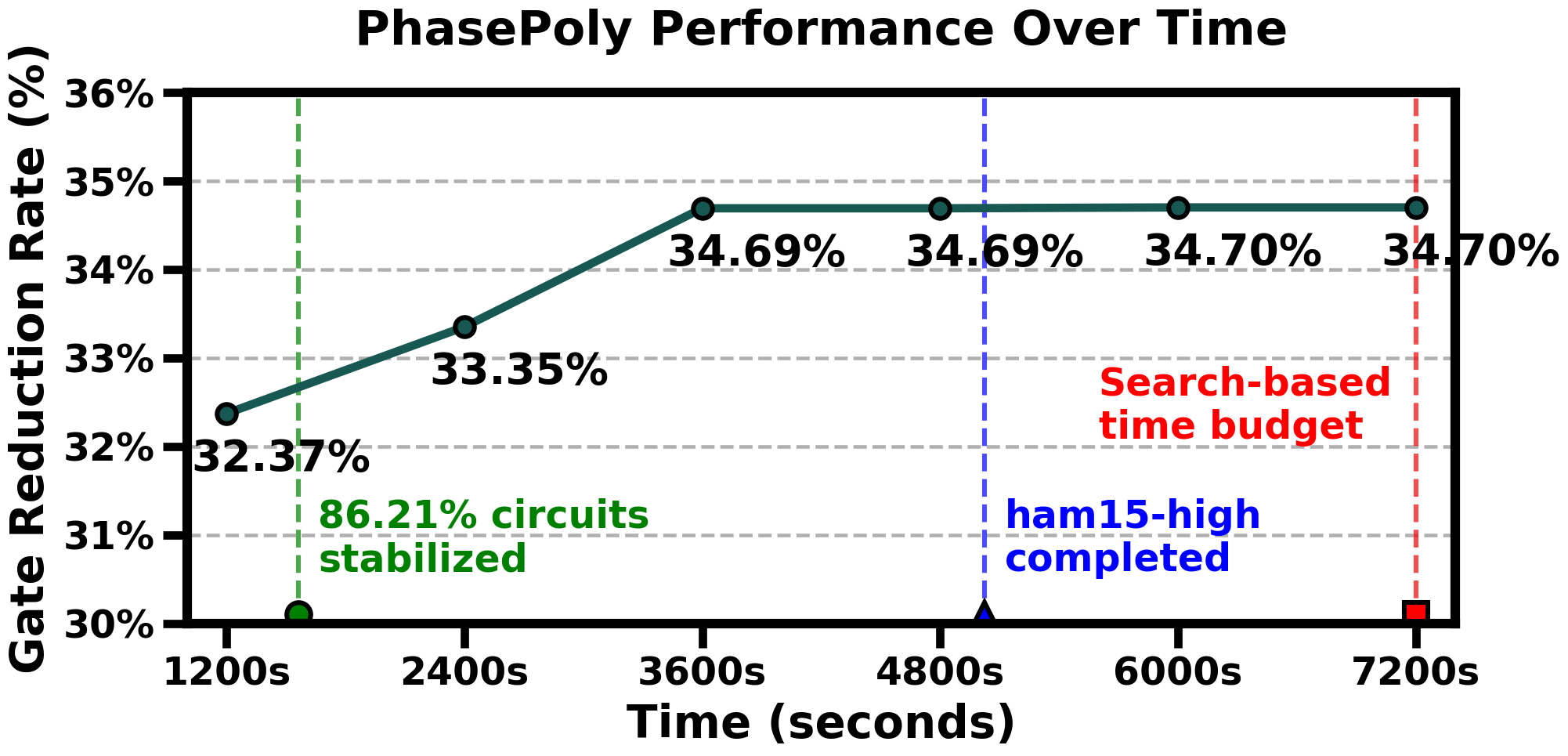}
        \caption{
        Optimization progress of \emph{PhasePoly} over time. 
        Average reduction reaches 32.37\% at 1{,}200\,s and converges near 34.7\% by 3{,}600--4{,}800\,s.
        }
        \vspace{-0.5\baselineskip}
        \label{fig:time_analysis}
    \end{figure}

\textbf{Compilation time.}
Fig.~\ref{fig:time_analysis} shows optimization progress under increasing runtime budgets using a deliberate over-optimization \emph{Incremental Block Merging} strategy. 
\emph{PhasePoly} reaches 32.37\% average gate reduction within 1{,}200\,s, and 86.21\% of benchmarks stabilize by 1{,}562\,s. 
Reductions further improve to 33.35\% at 2{,}400\,s and converge around 34.69\% by 3{,}600\,s. The slowest case, \texttt{ham15-high}, finishes in 5{,}025\,s, still below the 7{,}200\,s budget used for search-based subcircuit rewriting, while achieving substantially larger reductions.

\textbf{Parameter sensitivity.}
    We evaluate the sensitivity of \emph{PhasePoly}'s search parameters to:
    (1) priority-queue bound $Q$,
    (2) solution-pool size $P$, and
    (3) cross-block group size $G$.
    We denote settings as $(Q, P, G)$ and test the five largest circuits in our benchmark suite (12--28 qubits, 900--36{,}598 gates).

\begin{figure}[t!]
    \centering
    \vspace{-1.0\baselineskip}
        \includegraphics[width=0.24\textwidth]{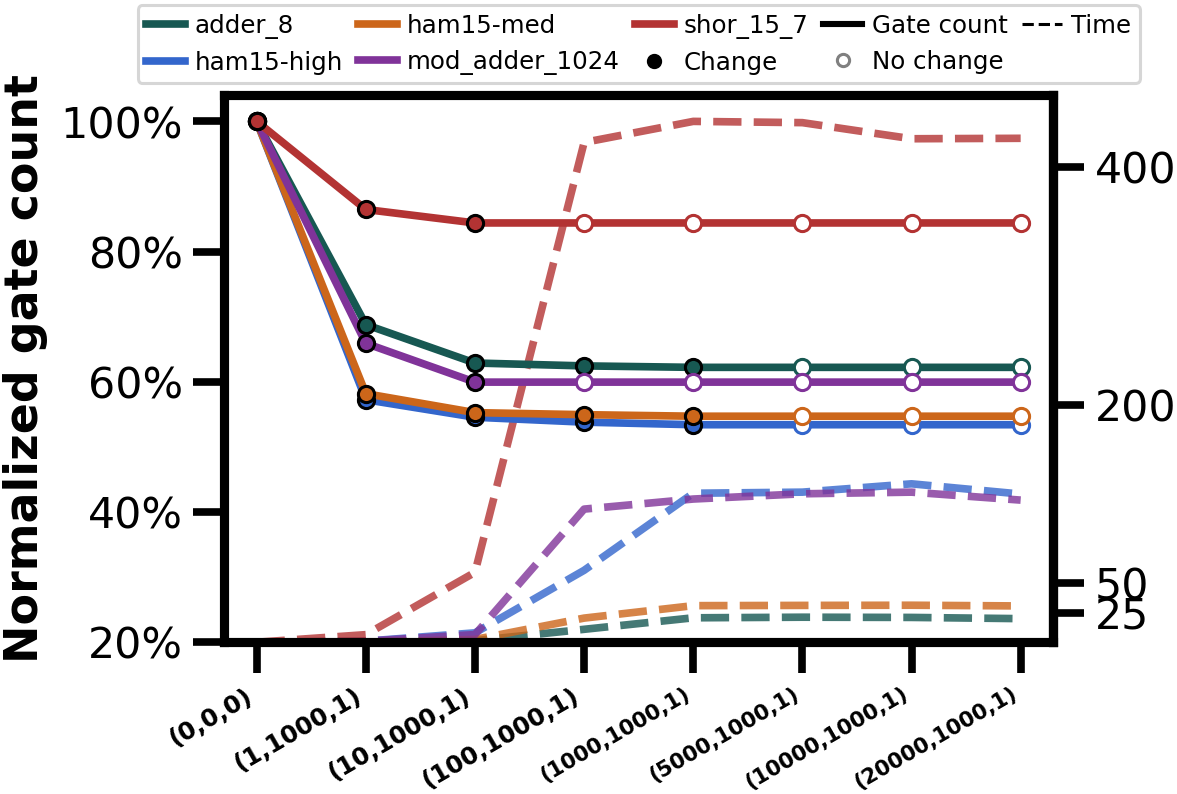}
        \includegraphics[width=0.24\textwidth]{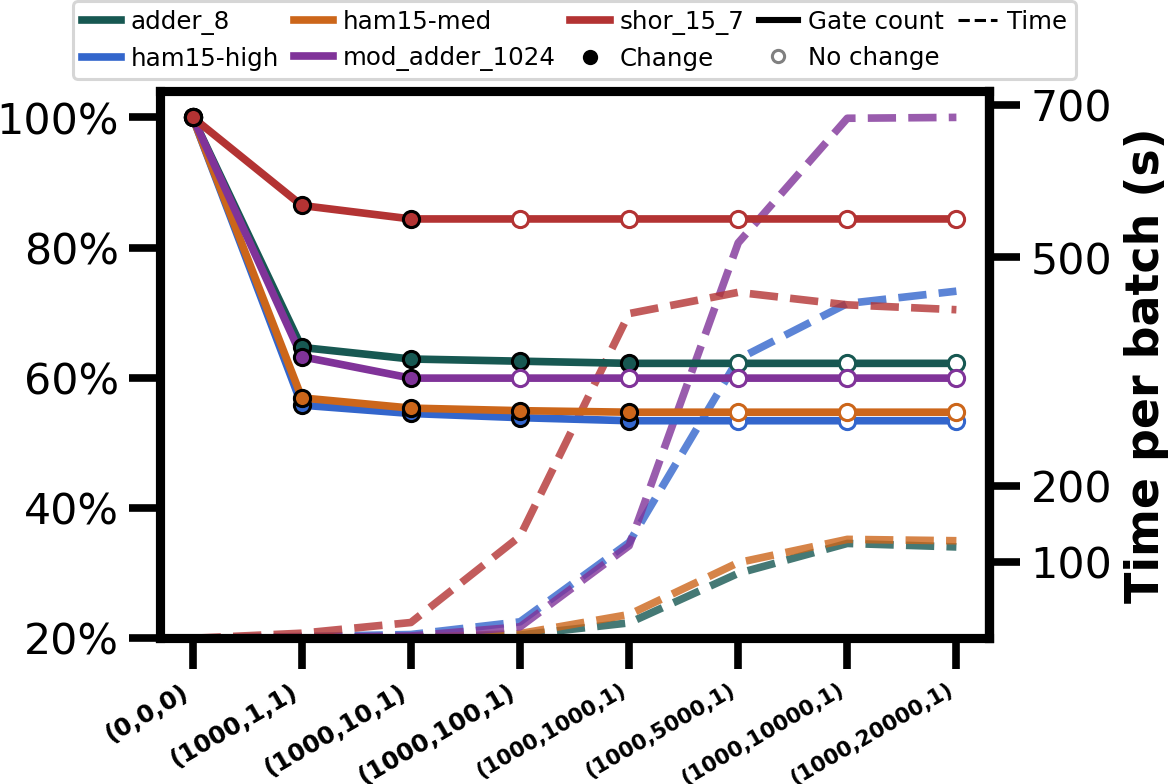}
        \caption{
            Sensitivity to $Q$ and $P$ with $G{=}1$ (no cross-block optimization). Each x-tick is $(Q, P, G)$. Left y-axis: normalized gate count; right y-axis: time needed (seconds).
            Solid: normalized reduction; dashed: runtime. Filled markers: improved quality; hollow markers: no change. Quality improves initially, then quickly saturates, while runtime continues to increase, especially with larger $P$.
        }
\vspace{-0.5\baselineskip}        
\label{fig:sensitivity_parameters_analysis_2}
    \end{figure}

\textbf{Queue/pool sizes: diminishing returns.}
With cross-block disabled ($G{=}1$), we vary $Q$ and $P$ from 1 to 20{,}000 separately. 
(i) When $Q$ is fixed at 1000, increasing $P$ improves quality only up to a moderate bound ($P\in[100,1000]$), after which reductions plateau while runtime increases (Fig.~\ref{fig:sensitivity_parameters_analysis_2} right). 
Similarly, with $P{=}1000$, quality also saturates at $Q\in[100,1000]$ (Fig.~\ref{fig:sensitivity_parameters_analysis_2} left). 
After that, although we increase the queue size to 20{,}000, the runtime grows slowly, which indicates that moderate settings ($Q\in[100,1000]$) are sufficient to reliably discover the top $\sim1000$ candidate solutions without requiring a larger search space.

\begin{figure}[t!]
    \centering
    \vspace{-1.0\baselineskip}
        \includegraphics[width=0.24\textwidth]{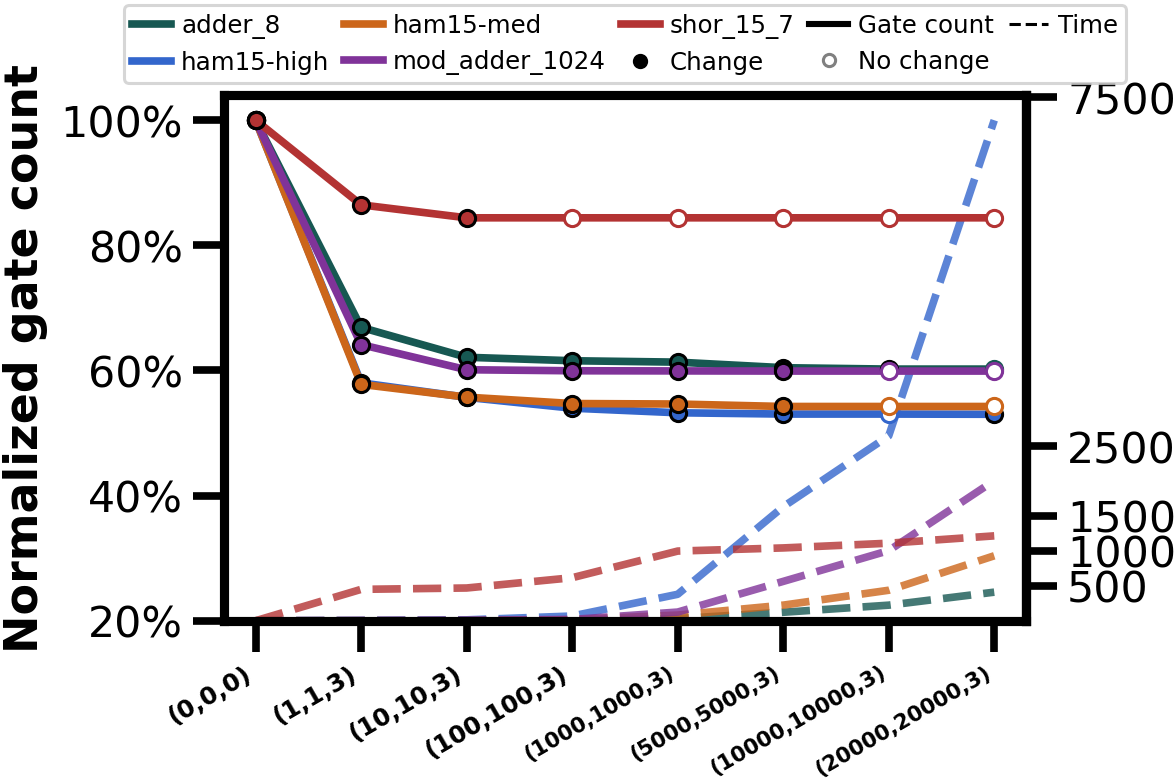}
        \includegraphics[width=0.24\textwidth]{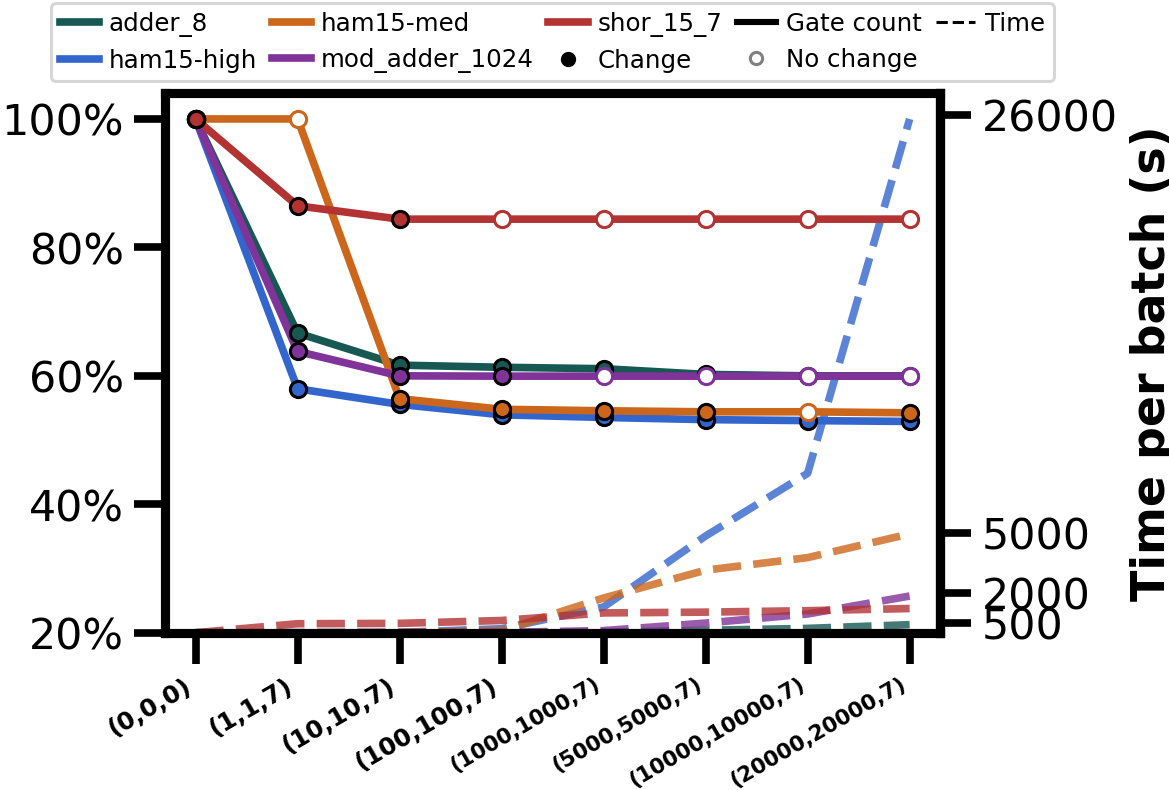}
        \caption{
Joint scaling of $(Q, P, G)$ for $G\in\{3,7\}$. Reductions saturate at moderate bounds (e.g., $Q{=}P{=}1000$),
while runtime grows rapidly for larger bounds. Extremely tight bounds can miss feasible cross-block solutions in rare cases. When $G\in\{1,5\}$, the trend is also consistent.}
\vspace{-0.5\baselineskip}
\label{fig:sensitivity_parameters_analysis}
    \end{figure}

\textbf{Joint parameter scaling and feasibility under cross-block optimization.}
Jointly scaling $(Q, P, G)$ confirms the same robustness trend. 
Fig.~\ref{fig:sensitivity_parameters_analysis} shows representative results for $G\in\{3,7\}$; results for $G\in\{1,5\}$ follow the same pattern and are omitted for readability. 
Across all group sizes, reductions saturate near $Q{=}P{=}1000$, while larger bounds provide only marginal quality improvement at substantially higher runtime. 
For example, increasing the solution pool from 1{,}000 to 20{,}000 increases runtime by nearly $20\times$ but yields only marginal additional reductions for \texttt{ham15-high} and \texttt{mod\_adder\_1024} across different group-size settings. 
When the bounds are extremely tight ($Q{=}P{=}1$), the bounded search may discard states needed to satisfy the rank-based correctness constraints, causing rare optimization failures; we observed this for \texttt{ham15-med} with $G\in\{5,7\}$.

    \vspace{-0.5\baselineskip}

    \begin{rqsummary}
\textbf{Q7 Summary:} 
\emph{PhasePoly} converges under moderate runtime budgets even under a deliberate over-optimization incremental block merging strategy. It is also robust to search parameters: moderate bounds
($Q{=}P{=}1000$) consistently achieve near-optimal reductions, while
larger search spaces mainly increase compilation time with diminishing
returns.
\end{rqsummary}
    \vspace{-0.5\baselineskip}

\section{Related Work}

\noindent\textbf{\textit{Manual Rule-based Optimization.}}
Many optimizers rely on manually written, equivalence-preserving rewrite rules~\cite{gadi_aleksandrowicz_2019_zenodo, smith2020open, sivarajah_2020_QuantumScienceTechnology, jang2023quixote}.
To ensure correctness, verification-oriented compilers and optimizers have been 
developed~\cite{amy2017verified,shi2019certiq,burgholzer2020advanced,
hietala2021POPL,arora2025local,liu2025popqc}, guaranteeing each applied rule is equivalence-preserving.

\noindent\textbf{\textit{Search-Based Subcircuit Rewriting.}}
Many optimizers rely on rule-based \emph{subcircuit rewriting}, where small patterns are matched and replaced.  
Systems such as Quanto~\cite{pointing2024QuantumScienceTechnology}, Quartz~\cite{xu2022PLDI}, and QUESO~\cite{xu2023PLDI} generate such rules automatically and apply them through global search, but their patterns are usually limited to small 3-qubit/6-gate regions, restricting long-range improvements.  
Reinforcement-learning approaches~\cite{fosel2021quantum,li2024OOPSLA, nagele2024optimizing} explore larger spaces but still depend on fixed rule sets and require costly pretraining.

\noindent\textbf{\textit{Phase Polynomial Optimization.}}
Prior work typically optimizes only the \emph{phase-parity} network~\cite{amy2018controlled,de2020architecture,nash2020quantum,vandaele2022QuantumScienceTechnology}, leaving the output-parity network to other passes.  
Nam \etal~\cite{nam2018npj} consider both but focus on 
random floating and merging of rotation gates.  
Other phase polynomial methods~\cite{amy2014polynomial,amy2019t,heyfron2019efficient,vandaele2025lower,ruiz2025quantum} mainly target $T$-count (often combined with higher-level techniques such as tensor-rank decomposition) rather than full CNOT/$R_z$ optimization.

\noindent\textbf{\textit{Unitary Synthesis and Hamiltonian Decomposition.}}
Unitary-synthesis approaches~\cite{doecode_58510,wu2020qgo,patel2022quest,kang2023modular,weiden2023improving,paradis2024synthetiq,weiden2024high,xu2024asplos,hao2026reducing}
optimize programs by synthesizing circuits for target unitaries. However,
they often rely on approximate equivalence, requiring explicit error
budgeting, and their scalability is limited.
Domain-specific decompositions, such as those for Hamiltonian
simulation~\cite{li+:asplos22,jin+:isca24,liu2024fermihedral,chen2025genesis, jang2026toward,zhou2024bosehedral},
achieve application- and hardware-driven improvements rather than general-purpose, exactly equivalent circuit optimization.

\section{Conclusion}

We revisit \emph{phase polynomial} optimization with \emph{PhasePoly}, which jointly optimizes the \emph{phase-parity} and \emph{output-parity} networks using a cross-block intermediate representation to enable long-range optimizations beyond single blocks. 
We advocate making phase polynomial optimization a standard component of the compilation pipeline.

\section*{Acknowledgments}

The authors thank Zirui Li, Minghao Guo, Jiakang Li, and Caitlin Chan (Rutgers University), Hanyu Wang (UCLA), and Mu-Te Lau (Northwestern University) for helpful discussions and feedback. 
This work was funded by Rutgers Research Council, the National Science Foundation (NSF), and the U.S. Department of Energy (DOE). 
In particular, Z.C., H.C., V.C., and E.Z. were supported by DOE Award DE-SC0025563, NSF Award CCF-2129872, NSF Award CCF-2529338, and a Rutgers Research Council Grant. 
This work was also funded by the National Research Foundation of Korea (NRF) under the project, “Creation of the Quantum Information Science R\&D Ecosystem Based on Human Resource” (RS-2023-00303229). 

\bibliographystyle{IEEEtran}
\bibliography{main}

\end{document}